\PassOptionsToPackage{dvipsnames}{xcolor}
\documentclass[11pt,letterpaper]{article}

\usepackage[numbers,sort&compress]{natbib}
\usepackage{authblk}
\usepackage[english]{babel}
\usepackage[T1]{fontenc} 

\usepackage{amsmath,amsfonts,amssymb,amsthm}
\usepackage{graphicx,graphbox}
\usepackage{float}

\usepackage[dvipsnames]{xcolor}
\usepackage[linktoc=all]{hyperref}
\hypersetup{colorlinks=true,linkcolor=RoyalBlue,
 anchorcolor = RoyalBlue,
 citecolor = RoyalBlue,
 filecolor = RoyalBlue,
 urlcolor = RoyalBlue}
\usepackage[left=1.2in,right=1.2in,top=1in,bottom=1in]{geometry}   
\usepackage{tocloft}

\usepackage{dsfont}
\usepackage{subcaption}
\usepackage[format=plain,font=small,textfont=it]{caption}
\usepackage{multirow} 
\usepackage{bookmark}
\usepackage{mathrsfs}
\usepackage{bm}
\usepackage{enumitem}

\newcommand{\ee}{\mathrm{e}}
\newcommand{\ii}{\mathrm{i}}

\renewcommand{\O}{\mathcal{O}}
\newcommand{\loc}{\mathrm{loc}}

\graphicspath{{img/}}

\begin{document}

\title{
\textbf{\LARGE \mbox{Typical entanglement entropy with charge conservation}}
}

\author{
Eugenio Bianchi,${}^{ab}\;$ 
Pietro Don\`a$\,{}^{c}\;$ and 
Erick Mui\~no$\,{}^{ab}$
}
\date{}

\maketitle

\begin{center}
\vspace{-3em}
{\footnotesize 
\href{mailto:ebianchi@psu.edu}{ebianchi@psu.edu}$\,,\;\;$
\href{mailto:pietro.dona@cpt.univ-mrs.fr}{pietro.dona@cpt.univ-mrs.fr}$\,,\;\;$
\href{mailto:erickmuino@psu.edu}{erickmuino@psu.edu}$\quad$\\[1em]
${}^{a}$ Department of Physics, The Pennsylvania State University, University Park, Pennsylvania 16802, USA\\[.3em]	
${}^{b}$ Institute for Gravitation and the Cosmos, The Pennsylvania State University,  Pennsylvania 16802, USA\\[.1em]
${}^{c}$ Aix-Marseille Univ, Universit\'e de Toulon, CNRS, CPT, Marseille, France
 }
\end{center}

\vspace{1em}

\begin{abstract}
We consider a many-body Hilbert space with a fixed global charge and show that the typical entanglement entropy of a subsystem, at the leading and subleading order in the thermodynamic limit, can be expressed in terms of a single quantity which represents the local thermal entropy at fixed charge density. We find a general formula which applies both to abelian $U(1)$ symmetry and non-abelian $SU(2)$ symmetry, including the case of a local Hilbert space which transforms under a general reducible representation of the symmetry group. We illustrate the general formula with model systems and discuss the relevance of the results as a probe of quantum chaos for physical Hamiltonians.
\end{abstract}

\noindent\rule{\textwidth}{0.1pt}

\setcounter{tocdepth}{2}
\tableofcontents

\noindent\rule{\textwidth}{0.1pt}


\section{Introduction}
\label{sec:intro}

In a seminal paper \cite{Page:1993df} Page showed that, in a $N$-qubit system with $N\gg 1$, the typical entanglement entropy of a $N_A$-qubit subsystem scales as $\log(2) N_A$ for $N_A<N/2$. The coefficient $\log(2)$ can be understood as the entropy of a single qubit at infinite temperature.\footnote{Throughout the paper, logarithms are natural logarithms, $\log(\ee)=1$.}

For a system of $N$ qubits at fixed $U(1)$ charge $m$, the typical entanglement entropy with the charge constraint \cite{Bianchi:2019stn} is found to scale as $\eta(s) N_A$, with $s=m/N$ and $\eta(s)=-\frac{1-2s}{2}\log(\frac{1-2s}{2})-\frac{1+2s}{2}\log(\frac{1+2s}{2})$, \cite{Bianchi:2021aui,Vidmar:2017pak}. In this paper we uncover the thermodynamic interpretation of the function $\eta(s)$ and show that it can be understood as the local thermal entropy of the one-body system at fixed average charge. We find a general formula which applies both to the case of $U(1)$ abelian charge and $SU(2)$ non-abelian charge \cite{Bianchi:2024aim}. The function $\eta(s)$ is fully determined by the structure of the local one-body Hilbert space $\mathcal{H}_\loc$ \cite{Yauk:2023wbu} and its decomposition into irreducible representations of the local symmetry group. It is given by the expression
\begin{equation}
\label{eq:eta-intro}
\eta(s)=\log(k)-\mathcal{D}(p_s\|p_\circledast)\,,
\end{equation}
where $k$ is the dimension of $\mathcal{H}_\loc$, the probability distribution $p_s$ is the thermal distribution at fixed average charge $s$, the probability distribution $p_\circledast$ is the thermal distribution at infinite temperature, and $\mathcal{D}(p_s\|p_\circledast)$ is the relative entropy, or Kullback-Leibler divergence. When the infinite-temperature distribution $p_\circledast$ is the uniform distribution $1/k$, the local entropy $\eta(s)$ reduces to the Shannon entropy of $p_s$.  To our knowledge, the general formula \eqref{eq:eta-intro} is new: 
it applies to both $U(1)$ and $SU(2)$ charges, to generic local Hilbert spaces, and it provides a thermodynamic interpretation of the leading order of the typical entanglement of a subsystem in terms of the local thermal entropy. Moreover, we show that the subleading terms in the typical entanglement entropy can be expressed in terms of the temperature and the local heat capacity associated with the local entropy $\eta(s)$. The role of  temperature and heat capacity in subleading orders of the entanglement entropy was previously discussed in \cite{Murthy:2019qvb} in the context of energy eigenstates of many-body chaotic Hamiltonians. Here we show that these quantities determine the typical entanglement entropy of random states at fixed charge, independently of the Hamiltonian. These results extend the known expressions for $U(1)$ \cite{Bianchi:2019stn,Bianchi:2021aui,Bianchi:2024aim,Vidmar:2017pak,Murthy:2019qvb,Murciano:2022lsw,Lau:2022hvc,Kliczkowski:2023qmp,Swietek:2023fka,Rodriguez-Nieva:2023err,Yauk:2023wbu,Jonay:2022cwg,Langlett:2024sxk,Ghosh:2024rvs,Langlett:2025fam,Medos:2026pkq} and $SU(2)$ charges \cite{Bianchi:2024aim,Patil:2023wdw,Patil:2025ump,Chakraborty:2025ziy,Wu:2026jxe,Yauk:2026quy}, including the orders $\O(N^{1/2})$ and $\O(N^0)$ in the typical entanglement entropy of random states, and allowing for a generic local Hilbert space.

\medskip

The results presented here are related to a substantial body of literature. The notion of Page curve for random states without constraints was first introduced in \cite{Page:1993df}. The exact formula for the average and variance of the entanglement entropy in the presence of constraints was derived in \cite{Bianchi:2019stn}. In the case of non-abelian symmetries, group-invariant subsystems at fixed charge can be defined in terms of subalgebras of observables \cite{Zanardi:2004zz,Petz:2007book,Balachandran:2013cq,Balachandran:2013hga}. The exact formula for the average and the variance of the entanglement entropy for these subsystems was derived in \cite{Bianchi:2024aim}. These results are relevant for 
non-abelian entanglement asymmetry in random states \cite{Russotto:2024pqg,Russotto:2025cpn}, quantum thermodynamics with non-commuting conserved charges \cite{Halpern:2016hlp,YungerHalpern:2019mvj,Majidy:2022kzx,Majidy:2023xhm,Noh:2025osn}, in the context of eigenstate thermalization \cite{Deutsch:1991msp,Srednicki:1994mfb,Rigol:2007juv,Noh:2022ijv,Murthy:2022dao,Patil:2026rst}, quantum many-body scars \cite{ODea:2020ooe,Moudgalya:2021xlu}, symmetry-resolved entanglement entropy for non-abelian groups  \cite{Goldstein:2017bua,Calabrese:2021wvi,Milekhin:2021lmq,Kusuki:2023bsp}, and quantum reference frames \cite{Garmier:2025soc, Spalvieri:2025uau,Araujo-Regado:2025ejs}. 
Recent work has investigated experimental probes of typical entanglement with charge conservation using a superconducting quantum processor \cite{Yang:2026fnq}.

\medskip

The paper is organized as follows: In Sec.~\ref{sec:many-body} we introduce the general structure of many-body systems and introduce the relation between the local entropy $\eta(s)$ and the asymptotics of the dimension of the Hilbert space at a fixed global charge. In Sec.~\ref{sec:subsystem} we discuss the decomposition of the many-body Hilbert space at fixed charge into subsystems, and determine the asymptotics of the dimensions of the subsystems. In Sec.~\ref{sec:EE-general} we derive the general formula for the average entanglement entropy and its variance, which then specialize to the case of $U(1)$ charge in Sec.~\ref{sec:U1} and $SU(2)$ charge in Sec.~\ref{sec:SU2}. Finally in Sec.~\ref{sec:discussion} we discuss the relevance of these results for the entanglement entropy in eigenstates of physical Hamiltonians.

\section{Many-body Hilbert space and global conserved charge}
\label{sec:many-body}
We introduce the general structure of many-body systems and derive the relation between the local entropy $\eta(s)$ and the asymptotic dimension of the Hilbert space at a fixed global charge.

\subsection{k-dimensional local system and global conserved charge}
We consider a many-body system consisting of $N$ bodies, each with a finite-dimensional Hilbert space $\mathcal{H}_{\loc}\simeq \mathbb{C}^k$ with dimension $k$, i.e., $\dim \mathcal{H}_{\loc}=k$. For $k=2$, each body is a \emph{qubit} and, for larger $k\in \mathbb{N}$, a \emph{qukit}. The Hilbert space of the system is the tensor product of local Hilbert spaces:
\begin{equation}
\label{eq:HN}
\mathcal{H}_N=\;\underbrace{
\mathcal{H}_{\loc}\otimes\cdots\otimes \mathcal{H}_{\loc}\,
}_{N}\,.
\end{equation}
We assume that each body carries a unitary (possibly reducible) 
representation of a group $G$. We focus here on the abelian group 
$G=U(1)$ and the non-abelian group $G=SU(2)$. The generalization 
to any semisimple Lie group will be presented in a forthcoming 
paper \cite{SU3}.

The Hilbert space of the many-body system decomposes into a direct sum over sectors of fixed charge $q$, i.e.,
\begin{equation}
\label{eq:HNdecomp}
\mathcal{H}_N=\bigoplus_{q}\big(\mathcal{H}_{\mathrm{irrep}}^{(q)}\otimes \mathcal{H}_N^{(q)}\big)\,.
\end{equation}
For $U(1)$, the charge is the magnetic number, $q\equiv m\in\mathbb{Z}$, and each irreducible representation is one-dimensional, $\dim \mathcal{H}_{\mathrm{irrep}}^{(q)}=1$. 
For $SU(2)$, the charge is the spin, $q\equiv j=0,\frac{1}{2},1,\frac{3}{2},\ldots$, and the dimension of the irreducible representation of spin $j$ is $\dim \mathcal{H}_{\mathrm{irrep}}^{(j)}=2j+1$.

We denote an orthonormal basis of the irreducible representation space $\mathcal{H}_{\mathrm{irrep}}^{(q)}$ by $|q,\mu\rangle$, where $\mu$ takes $\dim \mathcal{H}_{\mathrm{irrep}}^{(q)}$ distinct values and splits the degeneracy of the states with charge $q$. Specifically, for $U(1)$, there is no need for an additional quantum number; conventionally, $\mu$ is taken equal to the charge value, i.e., $\mu\equiv m$. For $SU(2)$, this extra quantum number can be taken as the eigenvalue of the $z$-component of the spin, i.e., the magnetic number $\mu\equiv m=-j,\ldots,+j$.

\bigskip

In the decomposition \eqref{eq:HNdecomp}, the Hilbert space $\mathcal{H}_N^{(q)}$ carries the internal degrees of freedom of the many-body system which are invariant under $G$-transformations at fixed charge $q$. This Hilbert space is defined as the intertwiner space for representations of the group $G$, \cite{Fulton:2004uyc,Hall2003}:
\begin{equation}
\label{eq:HNq-def}
\mathcal{H}_N^{(q)}=\mathrm{Inv}_G\big(\mathcal{H}_{\mathrm{irrep}}^{(q)*}\otimes \underbrace{
\mathcal{H}_{\loc}\otimes\cdots\otimes \mathcal{H}_{\loc}\,
}_{N} \big)\,,
\end{equation}
where $\mathcal{H}_{\mathrm{irrep}}^{(q)*}$ is the dual Hilbert space of $\mathcal{H}_{\mathrm{irrep}}^{(q)}$, with dual basis $\langle q,\mu|$. We refer to \cite{Bianchi:2024aim} for a discussion of their role in  quantum geometry \cite{Bianchi:2010gc,Bianchi:2011ub} and loop quantum gravity \cite{Rovelli:2014ssa,Ashtekar:2021kfp}.

\bigskip

The local one-body Hilbert space can also be decomposed as a direct sum over the local charge $q_{\loc}$:
\begin{equation}
\label{eq:H-loc}
\mathcal{H}_{\loc}=\bigoplus_{q_{\loc}}\big(\mathcal{H}^{(q_{\loc})}_{\mathrm{irrep}}\otimes \mathcal{H}_{\loc}^{(q_{\loc})}\big)\,,
\end{equation}
and its dimension is $k=\dim\mathcal{H}_{\loc}=\sum_{q_{\loc}} (\dim \mathcal{H}^{(q_{\loc})}_{\mathrm{irrep}})\, a_{q_{\loc}}$, where $a_{q_{\loc}}=\dim\mathcal{H}_{\loc}^{(q_{\loc})}$ is the multiplicity of the irreducible representation $q_{\loc}$ in $\mathcal{H}_{\loc}$.  
An orthonormal basis of $\mathcal{H}_{\loc}$ has the form $|q_\loc,\mu_\loc,\nu_\loc\rangle=|q_\loc,\mu_\loc\rangle |q_\loc,\nu_\loc\rangle$, with $\nu_\loc=1,\ldots,a_{q_\loc}$.
Specifically, for $U(1)$, the irreducible representations are one-dimensional, and the dimension of the local Hilbert space is simply $k=\dim\mathcal{H}_{\loc}=\sum_{m_{\loc}}a_{m_{\loc}}$ where $m_{\loc}$ is the $U(1)$ charge. For $SU(2)$, the irreducible representations have dimension $2j+1$, and the dimension of the local Hilbert space is $k=\dim\mathcal{H}_{\loc}=\sum_{j_{\loc}}(2j_{\loc}+1)\,a_{j_{\loc}}$.

\bigskip
 
The dimension of the Hilbert space $\mathcal{H}_N^{(q)}$ defined in \eqref{eq:HNq-def}, which carries the internal degrees of freedom of the many-body system at fixed charge, can be computed as an integral over the group of the characters of the representations, as shown for instance in App.~A of \cite{Bianchi:2024aim}. Specifically, we have
\begin{equation}
\label{eq:dimensionDq-characters}
D_q\equiv\dim \mathcal{H}_N^{(q)}=\;\int_G \overline{\chi^{(q)}(g)}\; \big(\chi_\loc(g)\big)^N\, d\mu(g)\,,
\end{equation}
where we have defined the character \cite{Fulton:2004uyc,Hall2003} of the local Hilbert space representation as the weighted sum over irreducible representations of the group,
\begin{equation}
\label{eq:chi-loc-def}
\chi_\loc(g)=
\sum_{q_{\loc}}  a_{q_{\loc}} \,\chi^{(q_{\loc})}(g)\,.
\end{equation}
The integral is taken with respect to the normalized Haar measure $d\mu(g)$ on the group. The explicit form of the characters depends on the group and the parametrization used for the group elements. The formula \eqref{eq:dimensionDq-characters} is exact and can be used to compute the dimension $D_q$ as shown in \cite{Bianchi:2024aim}. Our goal is to study the thermodynamic limit; in the following we focus only on the asymptotic properties of these dimensions.

\subsection{Hilbert space dimension and the local entropy at fixed charge density}
We investigate the many-body limit, $N\to \infty$, in the regime where the total charge $q$ is scaled extensively, i.e., $q=Ns$ with a fixed charge density $s$. This regime is commonly called the \emph{thermodynamic limit}.

\bigskip

The asymptotic behavior of the dimensions $D_q$ can be computed using Laplace's method to evaluate the integral \eqref{eq:dimensionDq-characters}. Rather than providing the technical details of the asymptotic analysis, which depend on the specific group and decomposition of the local Hilbert space, we present here only the final result for the asymptotic expansion of $D_q$ as a function of $s$:
\begin{equation}
\label{eq:dimensionDq-Asympt}
D(s)\equiv D_{Ns}=\;\alpha_0(s)\Big(1+\frac{\alpha_1(s)}{N}+\O(N^{-2}) \Big)\,\sqrt{\frac{-\eta''(s)}{2\pi\,N}}\;\ee^{N \eta(s)} \, .
\end{equation}
The function $\eta(s)$ has a general form which depends on the group and on the decomposition of the local Hilbert space into irreducible representations of charge $q_{\loc}$, as we discuss below. The function $\alpha_0(s)$ turns out to be completely determined by the group and by $\eta(s)$; specifically $\alpha_0(s)=1$ for $U(1)$ (as shown in Sec.~\ref{sec:U1}), and $\alpha_0(s)=1-\ee^{\eta'(s)}$ for $SU(2)$ (as shown in Sec.~\ref{sec:SU2}). Finally, as we will show in Sec.~\ref{sec:EE-general}, the asymptotic expressions for the typical entanglement entropy at order $\O(N^{-1})$ require that we take into account the term $\alpha_1(s)/N$ in \eqref{eq:dimensionDq-Asympt}, but the dependence on the specific value of $\alpha_1(s)$ cancels out in the final result. 

\bigskip

The function $\eta(s)$ has a fascinating interpretation in terms of statistical mechanics and relative entropy.
Consider a probability distribution $p_{q_{\loc}\mu_{\loc}}(\beta)$ over the local charge $q_{\loc}$ with magnetic number $\mu_{\loc}$. This distribution is required to maximize the Shannon entropy at fixed average magnetic number, i.e.,
\begin{equation}
\label{eq:beta-star}
\sum_{q_{\loc}\mu_{\loc}} \mu_{\loc}\;p_{q_{\loc}\mu_{\loc}}(\beta_*)\;=\;s\,.
\end{equation}
As usual \cite{Balian:1991}, the solution takes the Gibbs form 
\begin{equation}
\label{eq:p-beta}
p_{q_{\loc}\mu_{\loc}}(\beta)=
\frac{a_{q_{\loc}} \ee^{-\beta\mu_{\loc}}
}{\sum_{q'_{\loc}\mu'_{\loc}} a_{q'_{\loc}}\ee^{-\beta\mu'_{\loc}}}  \,,
\end{equation}
where the parameter $\beta$ plays the role of inverse \emph{temperature}, which is expressed as a function of the charge density $s$ by solving the condition \eqref{eq:beta-star}, i.e., $\beta_*=\beta_*(s)$. The probability distribution \eqref{eq:p-beta} is fully determined by the decomposition of the local Hilbert space \eqref{eq:H-loc} into irreducible representations of charge $q_{\loc}$, and it does not depend on the many-body limit or the scaling of the total charge. 

\bigskip

We express the function $\eta(s)$ in \eqref{eq:dimensionDq-Asympt} as:
\begin{equation}
\label{eq:eta-s-def}
\setlength{\fboxsep}{9pt}
\boxed{
\quad
\eta(s)\,=\,\log(k)\,-\sum_{q_{\loc}\mu_{\loc}}p_{q_{\loc}\mu_{\loc}}(\beta_*(s))\,\log\Big(\frac{p_{q_{\loc}\mu_{\loc}}(\beta_*(s))}{p_{q_{\loc}\mu_{\loc}}(0)}\Big)
\quad
}
\end{equation}
This expression shows that $\eta(s)$ is the \emph{local entropy}. The first term $\log(k)$ is the logarithm of the dimension of the local Hilbert space $\mathcal{H}_\loc$. The second term is minus the relative entropy, or Kullback-Leibler divergence $\mathcal{D}(p_s\|p_\circledast)$, of the probability distribution $p_s=p_{q_{\loc}\mu_{\loc}}(\beta_*(s))$ at finite temperature with respect to the same distribution at infinite temperature $p_\circledast=p_{q_{\loc}\mu_{\loc}}(0)=a_{q_{\loc}}/k$, where $k=\sum_{q_{\loc}\mu_{\loc}} a_{q_{\loc}}$. We note that, while the infinite temperature distribution is equiprobable over charge sectors, $p_\circledast$ is not simply $1/k$ because in general the multiplicities $a_{q_\loc}$ need not equal 1. An alternative description in terms of equiprobable distinguishable states $|q_{\loc},\mu_{\loc},\nu_\loc\rangle\in \mathcal{H}_\loc$ is possible, as done for instance in Ch.~3 of \cite{Balian:1991} where the Shannon entropy is discussed instead. By introducing a probability distribution $\tilde{p}_{q_{\loc}\mu_{\loc}\nu_\loc}(\beta)$ which is equiprobable over the auxiliary label $\nu_\loc=1,\ldots,a_{q_\loc}$ such that $p_{q_{\loc}\mu_{\loc}}(\beta)=\sum_{\nu_\loc} \tilde{p}_{q_{\loc}\mu_{\loc}\nu_\loc}(\beta)$, one finds that $\eta(s)$ is simply the Shannon entropy of $\tilde{p}_{q_{\loc}\mu_{\loc}\nu_\loc}(\beta)$. Therefore, the function $\eta(s)$ given by \eqref{eq:eta-s-def} and appearing in the asymptotic expression \eqref{eq:dimensionDq-Asympt} equals the entropy of the local thermal state in $\mathcal{H}_\loc$ at fixed average charge density $s$. 

\bigskip

At leading order, the asymptotic formula \eqref{eq:dimensionDq-Asympt} can be written as
\begin{equation}
\dim \mathcal{H}_N^{(Ns)}\,\approx\, k^{N}\,\ee^{-N\mathcal{D}(p_s\|p_\circledast)}\,.
\end{equation}
The constraint of fixed charge $q=Ns$ reduces the dimension $k^{N}$ of the many-body Hilbert space by a factor which scales exponentially in $N$ with rate given by the one-body, or local, relative entropy $\mathcal{D}(p_s\|p_\circledast)=\sum\, p_s \log (p_s/p_\circledast)$. 

\bigskip

The function $\eta(s)$ and the density $s$ have the thermodynamic properties of local entropy and \emph{energy density}. Specifically, the first derivative of \eqref{eq:eta-s-def} is the inverse \emph{temperature}
\begin{equation}
\label{eq:eta-d}
\beta_*(s) =\eta'(s)\,,
\end{equation}
and the second derivative is related to the \emph{local heat capacity}
\begin{equation}
\label{eq:c-def}
c_*(s)\equiv \left.\frac{\partial s}{\partial \beta^{-1}}\right|_{\beta_*}= -\left.\beta^2\frac{\partial s}{\partial \beta}\right|_{\beta_*}=
-\frac{\beta_*(s)^2}{\beta'_*(s)}=-\frac{\eta'(s)^2}{\eta''(s)}\,.
\end{equation}
The temperature $\beta_*^{-1}$ can be either positive or negative. If the temperature is positive, the dimension $D(s)$ of the Hilbert space increases when we increase the local charge $s$. Conversely, it decreases for negative temperature. However, the local heat capacity $c_*$ is always positive, which means that an increase in the charge density corresponds to an increase in temperature. The proof is immediate: from \eqref{eq:beta-star}, $-\partial s/\partial \beta$ equals the variance of the local magnetic number $\mu_\loc$, which is a positive quantity by definition. The positivity of $c_*(s)$ implies that $\eta(s)$ is a concave function, i.e., 
\begin{equation}
\eta''(s)<0\,.
\end{equation}
We will provide concrete examples of $\eta(s)$ for $U(1)$ and $SU(2)$ charges in Sec.~\ref{sec:U1} and \ref{sec:SU2}.

\section{Subsystem decomposition and subsystem charge distribution}
\label{sec:subsystem}

We discuss the decomposition into subsystems of the many-body Hilbert space at fixed charge, and determine the asymptotic dimensions of the subsystems.

\subsection{Subsystem decomposition}
The Hilbert space of the many-body system \eqref{eq:HN} decomposes as a tensor product over local subsystems. Observables that probe only a subset of $N_A$ bodies define a subsystem $A$ with Hilbert space $\mathcal{H}_A$ and a decomposition of the Hilbert space as a tensor product,
\begin{equation}
\label{eq:HAHB}
\mathcal{H}_N=\mathcal{H}_A\otimes \mathcal{H}_B\,,
\end{equation}
where 
\begin{equation}
\mathcal{H}_A=\;\underbrace{
\mathcal{H}_{\loc}\otimes\cdots\otimes \mathcal{H}_{\loc}\,
}_{N_A}\quad \text{and}\quad \mathcal{H}_B=\;\underbrace{
\mathcal{H}_{\loc}\otimes\cdots\otimes \mathcal{H}_{\loc}\,
}_{N_B}\,,
\end{equation}
where the complement of $A$ consists of $N_B=N-N_A$ bodies. 

We now consider the case of fixed charge $q$ corresponding to the decomposition \eqref{eq:HNdecomp} of the Hilbert space of the system. If we restrict to observables in the subsystem $A$ which are also \emph{invariant} under $G$-transformations, i.e., $G$-local observables \cite{Bianchi:2024aim}, then the Hilbert space $\mathcal{H}_N^{(q)}$ of the internal degrees of freedom of the system at fixed global charge $q$ decomposes as
\begin{equation}
\label{eq:HNq}
\mathcal{H}_N^{(q)}=\bigoplus_{q_A}\big(\mathcal{H}_{A}^{(q_A)}\otimes \mathcal{H}_{B}^{(q,q_A)}\big)\,,
\end{equation}
where $q_A$ is the charge of the subsystem $A$. Concretely, we can write
\begin{align}
\mathcal{H}_{A}^{(q_A)}=&\; \mathrm{Inv}_G\big(\mathcal{H}_{\mathrm{irrep}}^{(q_A)*}\otimes \underbrace{
\mathcal{H}_{\loc}\otimes\cdots\otimes \mathcal{H}_{\loc}\,
}_{N_A} \big)\,,\\
\mathcal{H}_{B}^{(q,q_A)}=&\; \mathrm{Inv}_G\big(\mathcal{H}_{\mathrm{irrep}}^{(q)*}\otimes\mathcal{H}_{\mathrm{irrep}}^{(q_A)}\otimes \underbrace{
\mathcal{H}_{\loc}\otimes\cdots\otimes \mathcal{H}_{\loc}\,
}_{N_B} \big)\,.
\end{align}
While the Hilbert space of the subsystem $A$ depends only on its charge $q_A$, the Hilbert space of the complement, $\mathcal{H}_{B}^{(q,q_A)}$, depends on both the system and subsystem charges $(q,q_A)$.

\bigskip

We introduce shorthand notation for the dimensions of these Hilbert spaces:
\begin{equation}
D_q=\dim \mathcal{H}_N^{(q)}\,,\quad d_{q_A}=\dim \mathcal{H}_{A}^{(q_A)}\,,\quad b_{q,q_A}=\dim \mathcal{H}_{B}^{(q,q_A)}\,,
\end{equation}
satisfying, by construction, the relation $D_q=\sum_{q_A}d_{q_A}b_{q,q_A}$. To obtain the exact expression for the dimensions, we can write them as the integral over the group of the characters of the representations involved, as we did in \eqref{eq:dimensionDq-characters}. Specifically, we have
\begin{align}
\label{eq:dimensionDq-characters2}
d_{q_A}&=\dim \mathcal{H}_{A}^{(q_A)}=\int_G \overline{\chi^{(q_A)}(g)}\big(\chi_\loc(g)\big)^{N_A}\, d\mu(g)\,,\\[.5em]
b_{q,q_A}&=\dim \mathcal{H}_{B}^{(q,q_A)}=\int_G \overline{\chi^{(q)}(g)} \,\chi^{(q_A)}(g)\big(\chi_\loc(g)\big)^{N_B}\, d\mu(g)\,,
\label{eq:dimensionDq-characters2-b}
\end{align}
where $\chi_\loc(g)$ is defined in \eqref{eq:chi-loc-def}.

\bigskip

We investigate the \emph{thermodynamic limit} $N\to \infty$, where the subsystem size $N_A$, the total charge $q$, and the subsystem charge $q_A$ scale extensively, i.e., we introduce the densities:
\begin{equation}
\label{eq:f-s-t-def}
f=\frac{N_A}{N}\,,\qquad s=\frac{q}{N}\,,\qquad t=\frac{q_A}{N_A}\,,
\end{equation}
that stay fixed in the limit $N\to \infty$. To keep the discussion as general as possible, instead of deriving explicit formulas (as done for specific examples in Sec.~\ref{sec:U1} and \ref{sec:SU2}), we write the asymptotic behavior of the dimensions in the implicit form
\begin{align}
\label{eq:Asympt-Ddb}
d(t)&\equiv d_{Ntf}=\;\alpha_0(t)\Big(1+\tfrac{\alpha_1(t)}{f N}+\O(N^{-2})\Big)\,\sqrt{\frac{-\eta''(t)}{2\pi\,f N\, }}\,\;\ee^{f N \eta(t)} \, ,
\\[.5em]
b(s,t)&\equiv b_{Ns,Ntf}=\Big(1+\tfrac{\gamma_1(\frac{s-f t}{1-f})-(1-f)^2\,\alpha_1(\frac{s-f t}{1-f})}{(1-f)fN}+\O(N^{-2})\Big)\,\sqrt{\frac{-\eta''(\frac{s-f t}{1-f})}{2\pi\,(1-f)N\,}}\,\;\ee^{(1-f)N \eta(\frac{s-f t}{1-f})} \, .\nonumber
\end{align}
The functions $\eta(s)$, $\alpha_0(s)$, and $\alpha_1(s)$ are defined in \eqref{eq:dimensionDq-Asympt}. The formula for $d(t)$ has the same form as the formula for $D(s)$. The formula for $b(s,t)$ is slightly different, because the integrand in \eqref{eq:dimensionDq-characters2-b} contains the product of two characters rather than a single one. In both cases, $U(1)$ and $SU(2)$, the leading contribution is a character integral of the same type as
\eqref{eq:dimensionDq-characters}, evaluated for the complement charge
$\frac{q-q_A}{N_B}=\frac{s-ft}{1-f}$, which explains the dependence on the same function $\eta$ and its argument in $b(s,t)$.

The function $\gamma_1(s)$ is determined by the normalization condition $D_q=\sum_{q_A} d_{q_A} b_{q,q_A}$ for the asymptotic expression. Sums over the subsystem charge $q_A$ can be evaluated asymptotically by converting them to an integral over the subsystem charge density $t$, using the Riemann sum approximation:
\begin{equation}
\label{eq:sum-int}
\sum_{q_A} \to \int f N \, dt \,.
\end{equation}
The relation \eqref{eq:HNq} between the Hilbert space of the system and of the subsystems $A$ and $B$ implies that
\begin{equation}
\int d(t)b(s,t)\,fN dt\;=\big(1+\O(N^{-2})\big) D(s)\,.
\end{equation}
This condition allows us to express $\gamma_1(s)$ in terms of $\eta(s)$ and $\alpha_0(s)$:
\begin{equation}
\textstyle \gamma_1(s)=\;\frac{(1-f)^2}{2}\frac{\alpha_0''(s)}{\alpha_0(s)\,\eta''(s)}+\frac{1-(1-f)f}{24}\Big(\frac{3\eta''''(s)}{\eta''(s)^2}-\frac{4\eta'''(s)^2}{\eta''(s)^3}\Big) \,.
\end{equation}

\subsection{Probability distribution for the subsystem charge density $t$}

With these definitions, we introduce the probability distribution function $\varrho_N(t)$, defined as
\begin{equation}
\label{eq:varrho-prob}
\varrho_N(t)\equiv \frac{d(t)\,b(s,t)}{D(s)} Nf\,.
\end{equation}
By construction, we have that the probability is normalized up to subleading corrections of order $\O(N^{-2})$,
\begin{equation}
\int \varrho_N(t) dt\,=\,1+\O(N^{-2})\,.
\end{equation}
The average of any function $F(t)$ which is independent of $N$ and is smooth in a neighborhood of $t=s$ (which is the stationary point of the probability distribution $\varrho_N(t)$) is
\begin{equation}
\label{eq:averageFcont}
\overline{F\,}(s)\,\equiv\,\int F(t)\,\varrho_N(t)\,dt \,=\,F(s)
+\left(\frac{\alpha_0'(s)}{\alpha_0(s)}F'(s)+\frac{1}{2}F''(s)\right)\frac{1-f}{(-\eta''(s))\,f\,}\frac{1}{N}
+\O(N^{-2})\,.
\end{equation}
The average is independent of $\alpha_1(t)$ at this order. Equivalently, the formula \eqref{eq:averageFcont} for the average can be written in terms of the Taylor series of $F(t)$ around the point $t=s$, using the moments of the local charge density $t$:
\begin{align}
\overline{\,(t-s)\,}\textstyle =\, \frac{\alpha_0'(s)}{\alpha_0(s)}\frac{1-f}{(-\eta''(s))\,f}\, \frac{1}{N}+\O(N^{-2})\,, \qquad
\overline{\,(t-s)^2\,}\textstyle =\,\frac{1-f}{(-\eta''(s))\,f}\, \frac{1}{N}+\O(N^{-2})\,,
\end{align}
while higher order moments are subleading, i.e., $\overline{\,(t-s)^k\,}=\O(N^{-2})$ for $k\geq 3$. On the other hand, if the function $F(t)$ is not smooth in a neighborhood of the stationary point $t=s$, i.e., a function of the form
\begin{equation}
F(t)=
\begin{cases}
\;F_-(t)\quad&t<s\,,\\[.2em]
\;F_+(t)&t>s\,,
\end{cases}
\end{equation}
then we have that the average is given by
\begin{equation}
\label{eq:averageFdisc}
\overline{F\,}(s)=\int F(t)\,\varrho_N(t)\,dt\;=\;\frac{F_+(s) + F_-(s)}{2}\;+\;
\tfrac{C_{1/2}(s)}{\sqrt{N}}+\tfrac{C_1(s)\vphantom{C_{1/2}}}{N\vphantom{\sqrt{N}}}
+\O(N^{-3/2})\,,
\end{equation}
with the coefficients
\begin{align}
C_{1/2}(s)=&\,\textstyle
\left(
\Big(\frac{(1-2 f)\,\eta'''(s)}{6 (1-f)\,\eta''(s)}
+ \frac{\alpha_0'(s)}{\alpha_0(s)}
\Big)(F_+(s) -F_-(s))
+ (F_+'(s) -F_-'(s))\right)
\sqrt{\frac{1-f}{2\pi\,(-\eta''(s))\,f}}\,,\\[.7em]
C_1(s)=&\,\textstyle
\left(
\frac{\alpha_0'(s)}{\alpha_0(s)}\frac{F_+'(s)+F_-'(s)}{2}
+\frac{F_+''(s)+F_-''(s)}{4}
\right)
\frac{1-f}{(-\eta''(s))\,f}\,.
\end{align} 
As concrete examples, we report the average of the functions $\theta(t-s)$ and $|t-s|$:
\begin{align}
\label{eq:averageAbs}
\textstyle \overline{\,\theta(t-s)\,}\,=\,\frac{1}{2}+\O(N^{-1/2})\,,\qquad\overline{\,|t-s|\,}\,=\,\sqrt{\frac{2(1-f)}{\pi(- \eta''(s))f}}\frac{1}{\sqrt{N}}+\O(N^{-1})\,.
\end{align}
In the next section, we use these expressions for computing the average entanglement entropy of a subsystem.


\section{Typical entanglement entropy at fixed charge}
\label{sec:EE-general}

We derive the general formula for the average entanglement entropy at fixed global charge and its variance using Laplace's method, and establish typicality in the thermodynamic limit.

\subsection{Average entanglement entropy: exact formula}
We consider a random pure state in the Hilbert space $\mathcal{H}_N^{(q)}$ with a fixed charge $q$, and its restriction to the $G$-invariant reduced state in the subsystem $A$ given by the decomposition \eqref{eq:HNq}.  For a non-abelian group $G$, the $G$-invariant entanglement entropy for this subsystem is defined in \cite{Bianchi:2024aim}. The average over random states of fixed charge is given by the exact formulas (98)--(99) in \cite{Bianchi:2024aim}, which can be rewritten as:
\begin{align}
\label{eq:entanglement_entropy_A}
\langle S_A\rangle_q\;=  \;\Psi(D_q+1)\;&-\sum_{q_A}\Big(\Psi\big(\!\max(d_{q_A},b_{q,q_A})+1\big)-\frac{1}{2\max(d_{q_A},b_{q,q_A})}\Big)\;\tfrac{d_{q_A}\, b_{q,q_A}}{D_q}\\[.5em]
&-\tfrac{1}{2}\sum_{q_A} \min\big(\tfrac{d_{q_A}}{b_{q,q_A}},\tfrac{b_{q,q_A}}{d_{q_A}}\big)\,\tfrac{d_{q_A}\, b_{q,q_A}}{D_q}  \,,
\label{eq:entanglement_entropy_A-min}
\end{align}
where $\Psi(x)=\Gamma'(x)/\Gamma(x)$ is the digamma function, i.e., the logarithmic derivative of  the gamma function $\Gamma(x)$. We are interested in the thermodynamic limit $N\to\infty$ of $\langle S_A\rangle_{N s}$ at fixed fraction $f$ and fixed charge density $s$, \eqref{eq:f-s-t-def}. We use the expressions \eqref{eq:dimensionDq-Asympt} and \eqref{eq:Asympt-Ddb} for the dimensions $D_q$, $d_{q_A}$, $b_{q,q_A}$, and approximate the sum over the local charge $q_A$ with an integral over $t$, up to order $\O(N^{-1})$, as done in \eqref{eq:sum-int}.

\subsection{Asymptotics and typicality}
We consider separately the three terms appearing in the expression \eqref{eq:entanglement_entropy_A}--\eqref{eq:entanglement_entropy_A-min}  for the average entanglement entropy $\langle S_A\rangle_q=Y_1+Y_2+Y_3$: 

\subsubsection{$Y_1$ term}
The first term in Eq.~\eqref{eq:entanglement_entropy_A} is $Y_1=\Psi(D_{q}+1)=\log(D(s))\,+\O(N^{-2})$, as follows from  the asymptotic expansion of the digamma function for $x\gg 1$,
\begin{equation}
\label{eq:Psi-asympt}
\Psi(x+1)-\frac{1}{2x}=\log x+\O(x^{-2})\,.
\end{equation}
Therefore, using \eqref{eq:dimensionDq-Asympt}, the first term has an asymptotic expansion 
\begin{equation}
\textstyle Y_1=\eta(s)N-\frac{1}{2}\log N+\log\big(\alpha_0(s)\big)+\log\Big(\sqrt{\frac{-\eta''(s)}{2\pi}}\,\Big)+\O(N^{-1})\,.
\end{equation}

\subsubsection{$Y_2$ term}
The second term $Y_2$ in Eq.~\eqref{eq:entanglement_entropy_A} reduces to the average 
\begin{equation}
Y_2=-\int \max\!\big(\log d(t), \log b(s,t)\big)\;\varrho_N(t)  dt\;\,+\O(N^{-1})\,, 
\end{equation}
where we have used \eqref{eq:Psi-asympt} for the digamma function. Moreover, using the asymptotics \eqref{eq:Asympt-Ddb} of the dimensions, we find the structure
\begin{equation}
\label{eq:maxdb}
\max\!\big(\log d(t), \log b(s,t)\big)=\,F(t)N-\tfrac{1}{2}\log N+G(t)+\O(N^{-1})\,.
\end{equation}
where the function $F(t)$ in a neighborhood of $t=s$ up to the order $\O((t-s)^2)$ is
\begin{align}
\label{eq:FY2}
F(t) = 
\begin{cases}
\;\textstyle (1-f)\,\eta(s)-f\eta'(s)\,(t-s)+\tfrac{1}{2}\tfrac{f^2}{1-f}\eta''(s)\,(t-s)^2\,+\,\O((t-s)^3)\;\;& f<\frac{1}{2}\,,\\[.5em]
\;\textstyle \tfrac{1}{2}\eta(s)+\tfrac{1}{2}|\eta'(s)|\,|t-s|+\tfrac{1}{4}\eta''(s)\,(t-s)^2\,+\,\O((t-s)^3)\quad& f=\frac{1}{2}\,, \\[.5em]
\;\textstyle  f\eta(s)+f\eta'(s)\,(t-s)+\tfrac{1}{2}f\eta''(s)\,(t-s)^2\,+\,\O((t-s)^3) \,& f>\frac{1}{2}\,,
\end{cases}
\end{align}
and the function $G(t)$ in a neighborhood of $t=s$ up to the order $\O((t-s)^0)$ is
\begin{align}
\label{eq:GY2}
G(t) = 
\begin{cases}
\;\textstyle   -\frac{\log(1-f)}{2}+\log\Big(\sqrt{\frac{-\eta''(s)}{2\pi}}\Big)\,+\,\O(t-s)\;\;& f<\frac{1}{2}\,,\\[.5em]
\;\textstyle -\frac{\log(1/2)}{2}+\log(\alpha_0(s))\,\theta\big((t-s)\eta'(s)\big)+\log\Big(\sqrt{\frac{-\eta''(s)}{2\pi}}\Big) \,+\,\O(t-s)\!& f=\frac{1}{2}\,, \\[.5em]
\;\textstyle -\frac{\log f}{2}\,+\,\log(\alpha_0(s))+\log\Big(\sqrt{\frac{-\eta''(s)}{2\pi}}\Big)\,+\,\O(t-s) & f>\frac{1}{2}\,.
\end{cases}
\end{align}
Using \eqref{eq:averageFcont}--\eqref{eq:averageFdisc}, we can compute the averages $\overline{F}(s)$ and $\overline{G}(s)$, and therefore the term  $Y_2$.\\
For $f<1/2$, we find
\begin{align}
\textstyle Y_2\big|_{f<\frac{1}{2}}=&\textstyle-(1-f)\eta(s) N+\frac{1}{2}\log N + \frac{\log(1-f) +f}{2}\\ 
&\textstyle- (1-f) \frac{\alpha_0'(s) \eta'(s)}{\alpha_0(s) \eta''(s)}-\log\Big(\sqrt{\frac{-\eta''(s)}{2\pi}}\Big) + \O(N^{-1})\,.\hspace{6em}
\end{align}
For $f=1/2$, using \eqref{eq:averageFdisc}, we find
\begin{align}
\textstyle Y_2\big|_{f=\frac{1}{2}}=&\textstyle - \frac{1}{2} \eta(s) N -\sqrt{\frac{\eta'(s)^2}{-2\pi \eta''(s)}} \sqrt{N}+\frac{1}{2}\log N+\frac{\log(1/2) +1/2}{2}\\[.5em]
&\textstyle -\frac{1}{2}\log(\alpha_0(s))-\log\Big(\sqrt{\frac{-\eta''(s)}{2\pi}}\Big)  \,+\, \O(N^{-1/2})\,.\hspace{6.6em}
\end{align}
Finally, for $f>1/2$, we find
\begin{align}
\textstyle Y_2\big|_{f>\frac{1}{2}}=&\textstyle -f \eta(s) N +\frac{1}{2}\log N+  \frac{\log(f) +1-f}{2}\\
&\textstyle +(1-f) \frac{\alpha_0'(s) \eta'(s)}{\alpha_0(s) \eta''(s)}-\log\big(\alpha_0(s)\big)-\log\Big(\sqrt{\frac{-\eta''(s)}{2\pi}}\Big) + \O(N^{-1})\,.
\end{align}

\subsubsection{$Y_3$ term}
The third term $Y_3$ in the expression \eqref{eq:entanglement_entropy_A-min} for $\langle S_A\rangle_q$ is bounded, $|Y_3|\leq \frac{1}{2}$, and is given by
\begin{equation}
\label{eq:Y3}
Y_3=-\tfrac{1}{2}\int \min\!\Big(\tfrac{d(t)}{b(s,t)},\tfrac{b(s,t)}{d(t)}\Big)\;\varrho_N(t) \, dt\;\,+\O(N^{-1})\,. 
\end{equation}
The minimum can be written as an exponential
\begin{equation}
\min\!\Big(\tfrac{d(t)}{b(s,t)},\tfrac{b(s,t)}{d(t)}\Big)=\ee^{-\left|\log \frac{d(t)}{b(s,t)}\right|}=(1+\O(N^{-1}))\,\tilde{h}(t)\,\ee^{- |\omega(t)| N}\,,
\end{equation} 
where $\tilde{h}(t)$ is a piecewise continuous function of $t$ of order $\O(N^0)$. The function $\omega(t)$ in the exponent is 
\begin{equation}
\omega(t)=(1-f)\, \eta(\tfrac{s-f t}{1-f})-f\,\eta(t)\,.
\end{equation}
Expanding around $t=s$, we find
\begin{equation}
\omega(t)\,=\,(1-2f)\eta(s)-2f\eta'(s)\,(t-s)+\O((t-s)^2)\,,
\end{equation}
which implies that  $Y_3$ is exponentially suppressed, unless $f=1/2$. Under this condition, we have $|\omega(t)|=|\eta'(s)|\,|t-s|+\O((t-s)^2)$. Using \eqref{eq:averageAbs}, we see that $\overline{\,|\omega(t)|\,}N\propto \sqrt{N}$, unless the subsystem fraction $f_{\circledast}$ and the global charge density $s_{\circledast}$ satisfy
\begin{equation}
f_{\circledast}=\tfrac{1}{2}\,,\quad\text{and}\quad \eta'(s_{\circledast})=0\,.
\end{equation}
Under these conditions, we can compute the value of $Y_3$, which is derived in App.~\ref{app:Y3},
\begin{equation}
\label{eq:Y3-general}
Y_3=-\tfrac{\alpha_0(s_\circledast)+\alpha_0(s_\circledast)^{-1}}{4}\,\delta_{s,s_\circledast}\,\delta_{f,\frac{1}{2}} \,+\,\O(N^{-1/2})\,.
\end{equation}

\subsubsection{Average entanglement entropy}
Combining the results above, the average entanglement entropy $\langle S_A\rangle_{q} $ for $q=Ns$ is
\begin{align}
\label{eq:entropyasymptotic}
\hspace{-.5em}
\langle S_A\rangle_{Ns} = 
\begin{cases}
\;\textstyle  f \eta(s) N + \frac{\log(1-f) +f}{2}  + \log(\alpha_0(s)) - (1-f) \frac{\alpha_0'(s) \eta'(s)}{\alpha_0(s) \eta''(s)} + \O(N^{-1}) \;\;& f<\frac{1}{2}\,,\\[1em]
\;\textstyle   \frac{1}{2} \eta(s) N -\sqrt{\frac{\eta'(s)^2}{-2\pi \eta''(s)}} \sqrt{N}+\frac{\log(1/2) +1/2}{2} +\frac{1}{2}\log(\alpha_0(s)) \,& f=\frac{1}{2}\,,\\[.5em]
\hspace{11.3em}-\frac{\alpha_0(s_\circledast)+\alpha_0(s_\circledast)^{-1}}{4}\delta_{s,s_\circledast} \,+\, \O(N^{-1/2})& \\[1em]
\;\textstyle  (1-f) \eta(s) N +  \frac{\log(f) +1-f}{2} +(1-f) \frac{\alpha_0'(s) \eta'(s)}{\alpha_0(s) \eta''(s)} + \O(N^{-1}) \,& f>\frac{1}{2}\,.
\end{cases}
\end{align}
The structure of $\langle S_A\rangle_{Ns}$ encompasses both the results of \cite{Bianchi:2021aui} for the systems with particle number conservation and of \cite{Bianchi:2024aim} for non-abelian $SU(2)$ charge and $G$-local subsystems, including also general local Hilbert spaces $\mathcal{H}_\loc$ \cite{Yauk:2023wbu}. The leading order of the average entanglement entropy is extensive up to half-system size $f\leq 1/2$, i.e., it is proportional to the number $N_A$ of bodies in the subsystem $A$. For larger subsystem fraction, $f>1/2$, the leading order is Page-like \cite{Page:1993df}, i.e., it can be obtained by the substitution $f\leftrightarrow 1-f$. The coefficient of the leading order term is the local entropy $\eta(s)$ defined in \eqref{eq:eta-s-def}.  At half-system size, $f=1/2$, a term of order $\O(\sqrt{N})$ appears \cite{Vidmar:2017pak}, with coefficient determined by the local heat capacity \eqref{eq:c-def} as identified in \cite{Murthy:2019qvb}. This term vanishes in the infinite temperature limit $s\to s_\circledast$. At order $\O(N^0)$, the universal term $\frac{\log(1-f) +f}{2}$ appears which was identified in \cite{Vidmar:2017pak} for eigenstates of a physical Hamiltonian with particle number conservation. This term was shown to be universal for random states with $U(1)$ charge in \cite{Yauk:2023wbu}, and now shown to be universal also for random states with $SU(2)$ charge, generalizing the result of \cite{Bianchi:2024aim} for spin-$1/2$ systems with a fixed $SU(2)$ charge (see \cite{Bianchi:2024aim} and \cite{Chakraborty:2025ziy} for a discussion of the relation between $K$-local and $G$-local subsystems). The terms that depend on $\alpha_0(s)$ are non-trivial only for $\alpha_0(s)\neq 1$, as it is the case for the non-abelian group $SU(2)$ where they result in an asymmetry under $f\leftrightarrow 1-f$ as identified in \cite{Bianchi:2024aim}. The half-system-size term $\delta_{s,s_\circledast}$ appears only at charge density $s=s_\circledast$ which corresponds to infinite temperature. As noticed in \cite{Yauk:2023wbu}, the $\delta_{s,s_\circledast}$ term and the $\O(\sqrt{N})$ term are mutually exclusive.

\subsubsection{Variance of the entanglement entropy}
The variance of the entanglement entropy for a random state of fixed charge $q$ is given by Eq.~(201) of \cite{Bianchi:2024aim}, (see also \cite{Bianchi:2019stn,Bianchi:2021aui}). In the thermodynamic limit, using the same methods described above, we find
\begin{equation}
\label{eq:variance-general}
(\Delta S_A)^2_{Ns}=\left(\Big(\sqrt{2\pi}\,f(1-f)-\tfrac{1}{\sqrt{2\pi}}\delta_{f,\frac{1}{2}}\Big)\frac{\eta'(s)^2}{\alpha_0(s)(-\eta''(s))^{3/2}}\,N^{3/2}+\O(N)\right)\ee^{-N\eta(s)}\,.
\end{equation}
For charge density $s$ corresponding to a non-vanishing local entropy, $\eta(s)\neq 0$,  the variance is exponentially suppressed in $N$. Therefore, in the thermodynamic limit, the average entanglement entropy $\langle S_A\rangle_{Ns}$ is typical: the probability $P(S_A)$ that a random state of charge $q$ has entropy $S_A$ is sharply peaked at the average. 

The proportionality coefficient in \eqref{eq:variance-general} depends on the product $c_*(s)^{3/2}\,|\beta_*(s)|^{-1}$ of the local heat capacity and the temperature. At this order, the formula shows that the variance is symmetric under the exchange $f\leftrightarrow 1-f$.

\section{Typical entanglement entropy at fixed $U(1)$ charge}
\label{sec:U1}
We derive the asymptotic formulas for general systems  at fixed $U(1)$ charge and discuss three model systems.

\subsection{Asymptotic formulas for general systems with a  $U(1)$ charge}
In this section we specialize the general formulas derived above to the case of a conserved $U(1)$ charge. In particular, we compute the functions $\eta(s)$ and $\alpha_0(s)$ appearing in the asymptotic expansion \eqref{eq:dimensionDq-Asympt}. We start from the expression \eqref{eq:dimensionDq-characters} for the dimension of the Hilbert space at fixed charge $q=Ns$. For $U(1)$, since the irreducible representations are one-dimensional, the character with charge $q=m$ is simply $\chi^{(m)}(g)=\ee^{\ii m \theta}$,  where $\theta$ parametrizes the group element $g=\ee^{\ii \theta}$. The Haar measure is $d\mu(g)=\frac{d\theta}{2\pi}$, and the character of the local Hilbert-space representation is
\begin{equation}
\chi_\loc(g)=\sum_{m_{\loc}} a_{m_{\loc}} \,\ee^{\ii m_{\loc} \theta}\,.
\end{equation}
With these definitions, the dimension of the Hilbert space at fixed charge $m=Ns$ can be written in the exact integral form
\begin{equation}
\label{eq:general-U1-dim}
D_{Ns}=\int_0^{2\pi} \frac{d\theta}{2\pi} \,\ee^{-\ii N s \theta}\Big(\sum_{m_{\loc}} a_{m_{\loc}} \,\ee^{\ii m_{\loc} \theta}\Big)^N = \int_0^{2\pi} \frac{d\theta}{2\pi} \,\ee^{N \tilde\eta(-\ii \theta,\,s)}\,,
\end{equation}
where we have defined the function
\begin{equation}
\tilde\eta(\beta,s)=\log\Big(\sum_{m_{\loc}} a_{m_{\loc}} \,\ee^{-\beta\,m_{\loc}}\Big) +  \beta\,s\,.
\end{equation}
The integral is dominated by the stationary point $\beta_*(s)$, determined by the condition 
\begin{equation}
\label{eq:saddleU1}
\partial_\beta \tilde\eta(\beta,s)\vert_{\beta_*(s)}=0 \quad \rightarrow \quad \frac{\sum_{m_{\loc}} m_{\loc} a_{m_{\loc}} \,\ee^{-\beta_*(s)\, m_{\loc}}}{\sum_{m_{\loc}} a_{m_{\loc}} \,\ee^{- \beta_*(s)\,m_{\loc}}} = s \,.
\end{equation}
Thus $\beta_*(s)$ is the inverse temperature of the thermal distribution \eqref{eq:p-beta} that maximizes the Shannon entropy at fixed average charge $s$. The stationary point $\beta_*(s)$ does not lie on the original integration contour, which is the imaginary axis, but it can be reached by deforming the contour in the complex plane.

Evaluating the function $\tilde\eta(\beta,s)$ at the stationary point $\beta_*(s)$ gives the function $\eta(s)$ appearing in \eqref{eq:dimensionDq-Asympt}. Completing the stationary-point calculation of $D_{Ns}$, we find
\begin{equation}
D_{Ns}\,=\,\frac{1}{2\pi}\sqrt{\frac{2\pi}{N\,\partial_\beta^2 \tilde\eta(\beta,s)\vert_{\beta_*(s)}}}\big(1+\O(1/N)\big)\,\ee^{N \tilde\eta(\beta_*(s),\,s)} \,.
\end{equation}
Defining $\eta(s)=\tilde\eta(\beta_*(s),\,s)$ and using the relation between the derivatives of $\eta(s)$ and those of $\tilde\eta(\beta,s)$ at the stationary point $\beta_*(s)$,
\begin{equation}
\eta'(s) = \beta_*(s) \quad \text{and}\quad \eta''(s) = -\frac{1}{\partial_\beta^2 \tilde\eta(\beta,s)\vert_{\beta_*(s)}}\,,
\end{equation}
we can express the asymptotic expansion of $D_{Ns}$ in the form \eqref{eq:dimensionDq-Asympt} as
\begin{equation}
D_{Ns} = \big(1+\O(N^{-1})\big)\, \sqrt{\frac{-\eta''(s)}{2\pi N}} \,\ee^{N \eta(s)}\,.
\end{equation}
We read off the functions $\eta(s)$ and $\alpha_0(s)$ from the asymptotic expansion of $D_{Ns}$ as
\begin{equation}
\eta(s) = \log\Big(\sum_{m_{\loc}} a_{m_{\loc}} \,\ee^{-\beta_*(s)\,m_{\loc}}\Big) + \beta_*(s)\,s \quad \text{and}\quad \alpha_0(s) = 1 \,.
\end{equation}
The coefficient $\alpha_1(s)$ can also be read off from a next-to-leading order calculation, but is not needed here. 
To make these formulas explicit, we now specify the decomposition of the local Hilbert space into irreducible representations of charge $m_{\loc}$, namely the coefficients $a_{m_{\loc}}$. 

\medskip

Let us summarize the generic features uncovered for the $U(1)$ charge case. The probability distribution over the local charge $m_{\loc}$ is given by
\begin{equation}
p_{m_{\loc}}(\beta)=\frac{a_{m_{\loc}} \,\ee^{-\beta m_{\loc}}}{\sum_{m_{\loc}} a_{m_{\loc}} \,\ee^{-\beta m_{\loc}}}\,.
\end{equation}
From the probability distribution $p_{m_{\loc}}(\beta)$, we define an inverse temperature $\beta_*(s)$ by fixing the average charge to be the system charge density $s$. Implicitly we have
\begin{equation}
\label{eq:beta-star-U1}
\sum_{m_{\loc}} m_{\loc}\;p_{m_{\loc}}(\beta_*)\;=\;s\,.
\end{equation}
The function $\eta(s)$ is given by $\log(k)$ minus the relative entropy of the probability distribution $p_{m_{\loc}}(\beta_*(s))$ with respect to the infinite-temperature distribution $p_{m_{\loc}}(0)=a_{m_{\loc}}/\sum_{m_{\loc}} a_{m_{\loc}}$,
\begin{equation}
\label{eq:eta-U1}
\eta(s)=\log(k)-\sum_{m_{\loc}}p_{m_{\loc}}(\beta_*(s))\log\Big(\frac{p_{m_{\loc}}(\beta_*(s))}{p_{m_{\loc}}(0)}\Big) \, .
\end{equation}
In terms of $\eta$, the inverse temperature $\beta_*(s)$ is given by the first derivative of $\eta(s)$, suggesting the interpretation of $\eta(s)$ as a thermodynamic potential:
\begin{equation}
\beta_*(s)=\eta'(s)\,.
\end{equation}
Similarly, we define the heat capacity from the second derivative of $\eta(s)$:
\begin{equation}
c_*(s)=\frac{\eta'(s)^2}{-\eta''(s)}\,.
\end{equation}
Finally, the coefficient $\alpha_0(s)$ appearing in the asymptotic expansion of $D_{Ns}$ is equal to $1$ for all $s$. This allows us to simplify the asymptotic expansion of the average entanglement entropy $\langle S_A\rangle_{Ns}$ computed in \eqref{eq:entropyasymptotic}. The average entanglement entropy with fixed $U(1)$ charge is:
\begin{equation}
\label{eq:entropy-U1}
\setlength{\fboxsep}{9pt}
\boxed{
\langle S_A\rangle_{Ns} =\begin{cases}
\;\textstyle  f \eta(s) N + \frac{\log(1-f) +f}{2} + \O(N^{-1}) \,& f<\frac{1}{2}\\[1em]
\;\textstyle   \frac{1}{2} \eta(s) N -\sqrt{\frac{c_*(s)}{2\pi}} \sqrt{N}+\frac{\log(1/2) +1/2}{2} -\frac{1}{2}\delta_{s,s_\circledast} + \O(N^{-{1/2}}) \,& f=\frac{1}{2}\\[1em]
\;\textstyle  (1-f) \eta(s) N +  \frac{\log(f) +1-f}{2}  + \O(N^{-1}) \,& f>\frac{1}{2}
\end{cases}
}
\end{equation}
The variance of the entanglement entropy with a fixed $U(1)$ charge is
\begin{equation}
\label{eq:variance-U1}
(\Delta S_A)^2_{Ns}=\left(\Big(\sqrt{2\pi}\,f(1-f)-\tfrac{1}{\sqrt{2\pi}}\delta_{f,\frac{1}{2}}\Big)\frac{c_*(s)^{3/2}}{|\beta_*(s)|}\,N^{3/2}+\O(N)\right)\ee^{-N\eta(s)}\,,
\end{equation}
which is exponentially suppressed in $N$ for any charge density $s$ such that $\eta(s)\neq 0$. Therefore, the average entanglement entropy $\langle S_A\rangle_{Ns}$ is typical in the thermodynamic limit. 

\medskip

These results match the ones previously found in \cite{Bianchi:2021aui} and \cite{Yauk:2023wbu}, providing a check of the general formulas \eqref{eq:entropyasymptotic} and \eqref{eq:variance-general}, together with a thermodynamic interpretation of the function $\eta(s)$ and its derivatives. We consider below the examples of a local qubit Hilbert space and of two-species hardcore bosons with a $U(1)$ charge, and show that they reproduce known results in the literature 
\cite{Bianchi:2019stn,Bianchi:2021aui,Bianchi:2024aim,Vidmar:2017pak,Murthy:2019qvb,Murciano:2022lsw,Lau:2022hvc,Kliczkowski:2023qmp,Swietek:2023fka,Rodriguez-Nieva:2023err,Yauk:2023wbu,Jonay:2022cwg,Langlett:2024sxk,Ghosh:2024rvs,Langlett:2025fam,Medos:2026pkq}.

\subsection{Example: $N$ qubits at fixed $U(1)$ charge (paramagnet)}

We consider the case of $N$ qubits. Each qubit Hilbert space carries the local $U(1)$ charges $m_{\loc}=\pm 1/2$, corresponding to the local Hilbert space decomposition 
\begin{equation}
\mathcal{H}_{\loc}=\mathcal{H}_{\mathrm{irrep}}^{(-1/2)}\oplus \mathcal{H}_{\mathrm{irrep}}^{(+1/2)}\,,
\end{equation}
with $a_{-1/2}=a_{+1/2}=1$. We compute the average entanglement entropy of this $N$-body system at fixed global $U(1)$ charge $m=Ns$ and fixed subsystem fraction $f$. The spectrum of the physical Hamiltonian of a paramagnetic system decomposes into $U(1)$ charge sectors of fixed magnetization \cite{Bianchi:2019stn,Balian:1991} and has the same structure as the one discussed here, with the charge $m$ proportional to the magnetization.

The local entropy $\eta(s)$ is defined in terms of the probability distribution 
\begin{equation}
  p_{m_{\loc}}(\beta)=\frac{\ee^{-\beta m_{\loc}}}{\ee^{-\beta/2}+\ee^{\beta/2}}\,,
\end{equation}
where the inverse temperature $\beta$ is fixed by the condition \eqref{eq:saddleU1}, which gives
\begin{equation}
  \frac{\tfrac{1}{2}\ee^{-\beta_*(s)/2}-\tfrac{1}{2}\ee^{\beta_*(s)/2}}{\ee^{-\beta_*(s)/2}+\ee^{\beta_*(s)/2}}= s\,, \quad \rightarrow \quad \beta_*(s) = -2\mathrm{arctanh}(2s)\,.
\end{equation}
The probabilities $p_{m_{\loc}}(\beta_*(s))$ in terms of $s$ are
\begin{equation*}
p_{-1/2}(s)=\frac{1-2s}{2},\qquad p_{+1/2}(s)=\frac{1+2s}{2}\,.
\end{equation*}
The function $\eta(s)$ can be computed explicitly as $-p_{-1/2}\log p_{-1/2}-p_{+1/2}\log p_{+1/2}$ and takes the form
\begin{equation}
\eta_{{U(1)-\mathrm{qubit}}}(s)\,\equiv\,-\frac{1-2s}{2}\log\Big(\frac{1-2s}{2}\Big)-\frac{1+2s}{2}\log\Big(\frac{1+2s}{2}\Big)\ .
\label{eq:eta-U1-qubit}
\end{equation}
Substituting this expression into \eqref{eq:entropyasymptotic} gives the entanglement entropy at fixed charge. This result reproduces the known results for the typical entanglement entropy of $N$ qubits with a fixed total abelian charge $q=Ns$ \cite{Bianchi:2019stn,Murciano:2022lsw,Yauk:2023wbu,Bianchi:2024aim}. The normalization $s=q/N$ differs from the convention $\tilde{s}=2q/N$ common in the literature, giving factors of $2s$ instead of $s$ in \eqref{eq:eta-U1-qubit}.

From $\eta(s)$ we can compute the temperature and the local heat capacity:
\begin{equation}
\label{eq:beta-c-U1-qubit}
\beta_*(s)=\log\Big(\frac{1 - 2 s}{1 + 2 s}\Big) \, , \qquad
c_*(s)=\big(\tfrac{1}{4}-s^2\big)\, \Big(\log\frac{1 - 2 s}{1 + 2 s}\,\Big)^2 \, .
\end{equation} 

\subsection{Example: $N$ qutrits at fixed $U(1)$ charge (softcore bosons)}
We consider the case of $N$ qutrits, $k=3$. Each qutrit Hilbert space carries the local $U(1)$ charges $m_{\loc}=-1,0,1$, corresponding to the local Hilbert space decomposition 
\begin{equation}
\mathcal{H}_{\loc}=\mathcal{H}_{\mathrm{irrep}}^{(-1)}\oplus\mathcal{H}_{\mathrm{irrep}}^{(0)}\oplus \mathcal{H}_{\mathrm{irrep}}^{(+1)}\,,
\end{equation}
with $a_{-1}=a_{0}=a_{+1}=1$. We compute the average entanglement entropy of this $N$-body system at fixed global $U(1)$ charge $m=Ns$ and fixed subsystem fraction $f$. The spectrum of the physical Hamiltonian of soft-core bosons decomposes into sectors of fixed number of particles \cite{Yauk:2023wbu,Patil:2026rst,Medos:2026pkq} and has the same structure as the one discussed here, with the global charge $m=B-N$ where $B$ is the number of particles. 

The local entropy $\eta(s)$ is defined in terms of the probability distribution 
\begin{equation}
  p_{m_{\loc}}(\beta)=\frac{\ee^{-\beta m_{\loc}}}{\ee^{-\beta}+1+\ee^{\beta}}\, ,
\end{equation}
where the inverse temperature $\beta$ is fixed by the condition \eqref{eq:saddleU1}, which gives
\begin{equation}
  \frac{\ee^{-\beta_*(s)}-\ee^{\beta_*(s)}}{\ee^{-\beta_*(s)}+1+\ee^{\beta_*(s)}}= s \, , \quad \rightarrow \quad \beta_*(s) = \log\Big(\frac{-s + \sqrt{4 - 3 s^2}}{2 (1 + s)}\Big)\,.
\end{equation}
The probabilities $p_{m_{\loc}}(\beta_*(s))$ in terms of $s$ are
\begin{equation*}
p_{-1}(s)=\frac{4 - 3 s - \sqrt{4 - 3 s^2}}{6}\, , \quad
p_{0}(s)=\frac{-1 + \sqrt{4 - 3 s^2}}{3}\, , \quad
p_{+1}(s)=\frac{4 + 3 s - \sqrt{4 - 3 s^2}}{6}\, .
\end{equation*}
With these definitions, the function $\eta(s)$ can be computed explicitly as $-p_{-1}\log p_{-1}-p_0\log p_0-p_{+1}\log p_{+1}$ and takes the form
\begin{equation}
\begin{aligned}
\eta_{{U(1)-\mathrm{qutrit}}}(s)\,\equiv&-\frac{4 - 3 s - \sqrt{4 - 3 s^2}}{6}\log\big(\frac{4 - 3 s - \sqrt{4 - 3 s^2}}{6}\big)\\
&-\frac{-1 + \sqrt{4 - 3 s^2}}{3}\log\big(\frac{-1 + \sqrt{4 - 3 s^2}}{3}\big)\\
&-\frac{4 + 3 s - \sqrt{4 - 3 s^2}}{6}\log\big(\frac{4 + 3 s - \sqrt{4 - 3 s^2}}{6}\big)\ ,
\end{aligned}
\label{eq:eta-U1-qutrit}
\end{equation}
which is equivalent to the function reported in \cite{Yauk:2023wbu,Patil:2026rst,Medos:2026pkq}, (after shifting the charge by $1$ due to different conventions). 

From $\eta(s)$ we can compute the temperature and the local heat capacity:
\begin{align}
\beta_*(s)&=\log\Big(\frac{-s + \sqrt{4 - 3 s^2}}{2 (1 + s)}\Big)\, , \\[.5em]
c_*(s)&= \frac{\sqrt{4-3 s^2}\big(\sqrt{4-3 s^2}-1\big)}{12} \left(\log\frac{4+3 s-\sqrt{4-3 s^2}}{4-3s-\sqrt{4-3 s^2}}\right)^2 \, .\label{eq:beta-c-U1-qutrit}
\end{align} 

\subsection{Example: $N$ qutrits at fixed $U(1)$ charge (two-species hardcore bosons)}
We consider again the case of $N$ qutrits, $k=3$. Each qutrit Hilbert space is now assumed to carry the local $U(1)$ charges $m_{\loc}=0,+1$ with multiplicities $a_0=1$ and $a_{+1}=2$, corresponding to the local Hilbert space decomposition 
\begin{equation}
\mathcal{H}_{\loc}=\mathcal{H}_{\mathrm{irrep}}^{(0)}\oplus \mathcal{H}_{\mathrm{irrep}}^{(+1)}\oplus \mathcal{H}_{\mathrm{irrep}}^{(+1)}\,.
\end{equation}
We compute the average entanglement entropy of this $N$-body system at fixed global $U(1)$ charge $m=Ns$ and fixed subsystem fraction $f$. The spectrum of the physical Hamiltonian of a system of two-species hardcore bosons decomposes into sectors of fixed number of particles \cite{Medos:2026pkq,Yauk:2023wbu} and has the same structure as the one discussed here, with the global charge $m=B_X+B_Y$ where $B_X$ and $B_Y$ are the number of particles of species $X$ and $Y$.

The local entropy $\eta(s)$ function is defined in terms of the probability distribution
\begin{equation}
  p_{m_{\loc}}(\beta)=\frac{a_{m_{\loc}}\ee^{-\beta m_{\loc}}}{1+2\,\ee^{-\beta}}\,.
\end{equation}
The inverse temperature $\beta$ is again fixed by the condition \eqref{eq:saddleU1}, which now gives
\begin{equation}
  \frac{2\ee^{-\beta_*(s)}}{1+2\,\ee^{-\beta_*(s)}}=s\,, \quad \rightarrow \quad \beta_*(s) = \log 2 + \log\frac{1 - s}{s}\,.
\end{equation}
The probabilities $p_{m_{\loc}}(\beta_*(s))$ in terms of $s$ are then
\begin{equation}
p_{0}(\beta_*(s))=1-s\, , \quad  p_{+1}(\beta_*(s))=s\,,
\end{equation}
and the probabilities at infinite temperature
\begin{equation}
p_{0}(0)=\tfrac{1}{3}\, , \quad  p_{+1}(0)=\tfrac{2}{3}\,,
\end{equation}
From the probabilities and the formula \eqref{eq:eta-U1} for the local entropy $\eta(s)$ we find
\begin{equation}
\eta_{{U(1)}-\mathrm{2bosons}}(s)\,\equiv\,\log(3)-(1 - s) \log\big(\tfrac{1 - s}{1/3}\big) - s \log\big(\tfrac{s}{2/3}\big)\, .
\label{eq:eta-U1-2bosons}
\end{equation}
The average entanglement entropy at fixed global charge is obtained by plugging this function into the general formulas \eqref{eq:entropyasymptotic}. This expression reproduces the known results for the typical entanglement entropy of two-species hardcore bosons at fixed number of particles $q=Ns$ \cite{Yauk:2023wbu}, providing a thermodynamic interpretation of the function $\eta(s)$.

From the local entropy $\eta(s)$ we can compute the temperature and the local heat capacity:
\begin{equation}
\label{eq:beta-c-U1-boson}
\beta_*(s)=\log\Big(\frac{2 - 2s}{s}\Big) \, , \qquad
c_*(s)=s(1-s) \, \Big(\log\frac{2 - 2s}{s}\Big)^2 \, .
\end{equation} 
The system has infinite temperature $\beta_*(s_\circledast)=0$ at the local charge density
\begin{equation}
s_\circledast=\frac{2}{3}\,,
\end{equation}
where the $\O(1)$ term \eqref{eq:Y3-general} appears, $Y_3=-\tfrac{1}{2}\,\delta_{s,\frac{2}{3}}\,\delta_{f,\frac{1}{2}}$.

\begin{figure}[t]
\centering
\includegraphics[width=.32\textwidth]{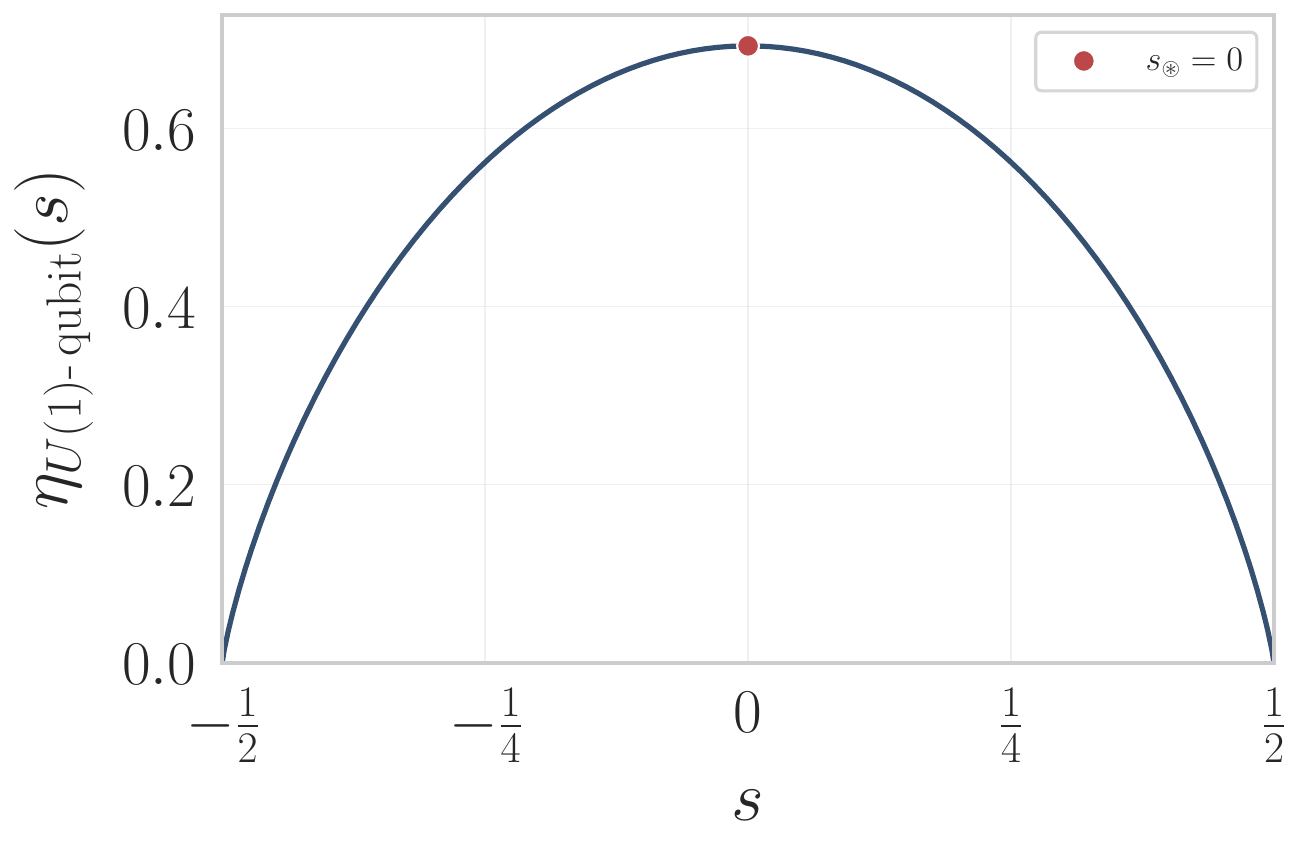}
\includegraphics[width=.32\textwidth]{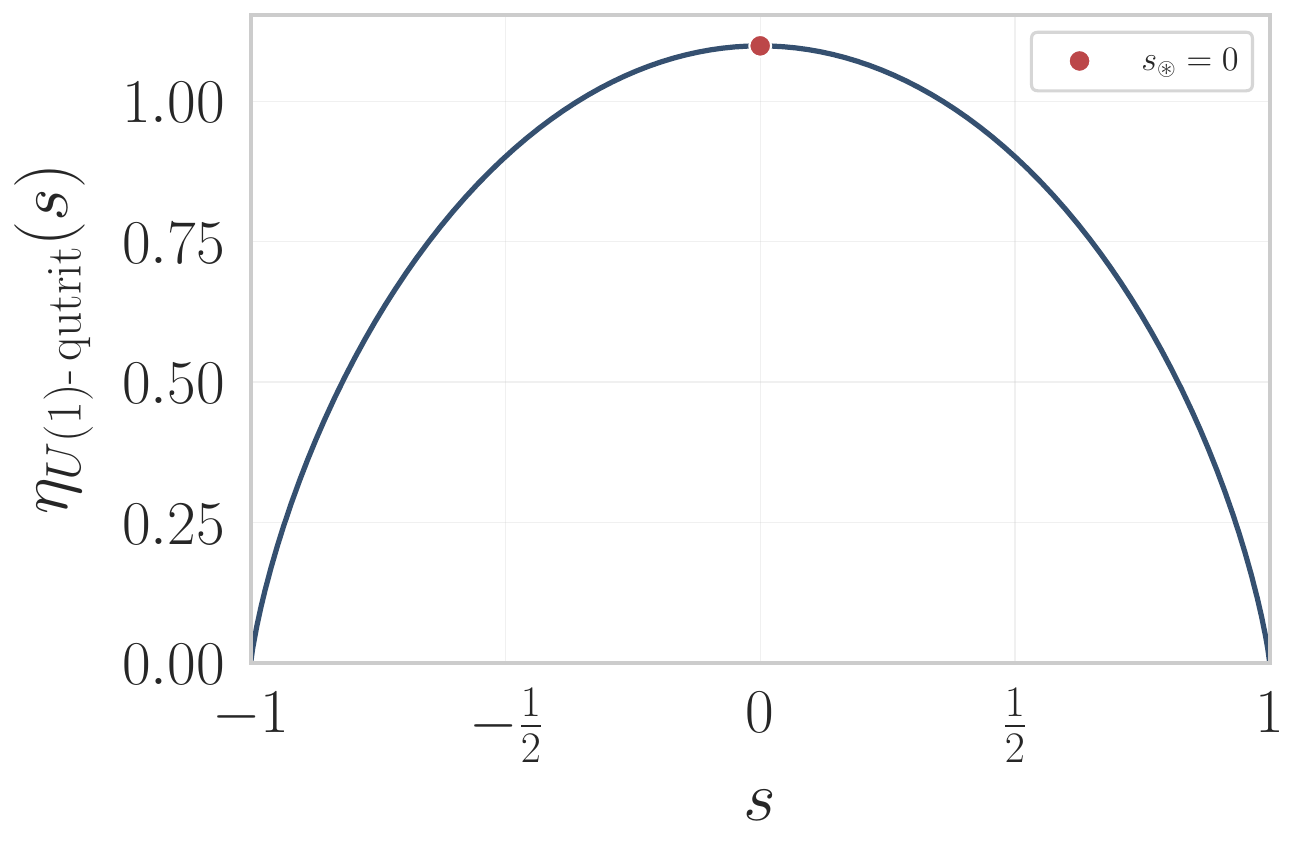}
\includegraphics[width=.32\textwidth]{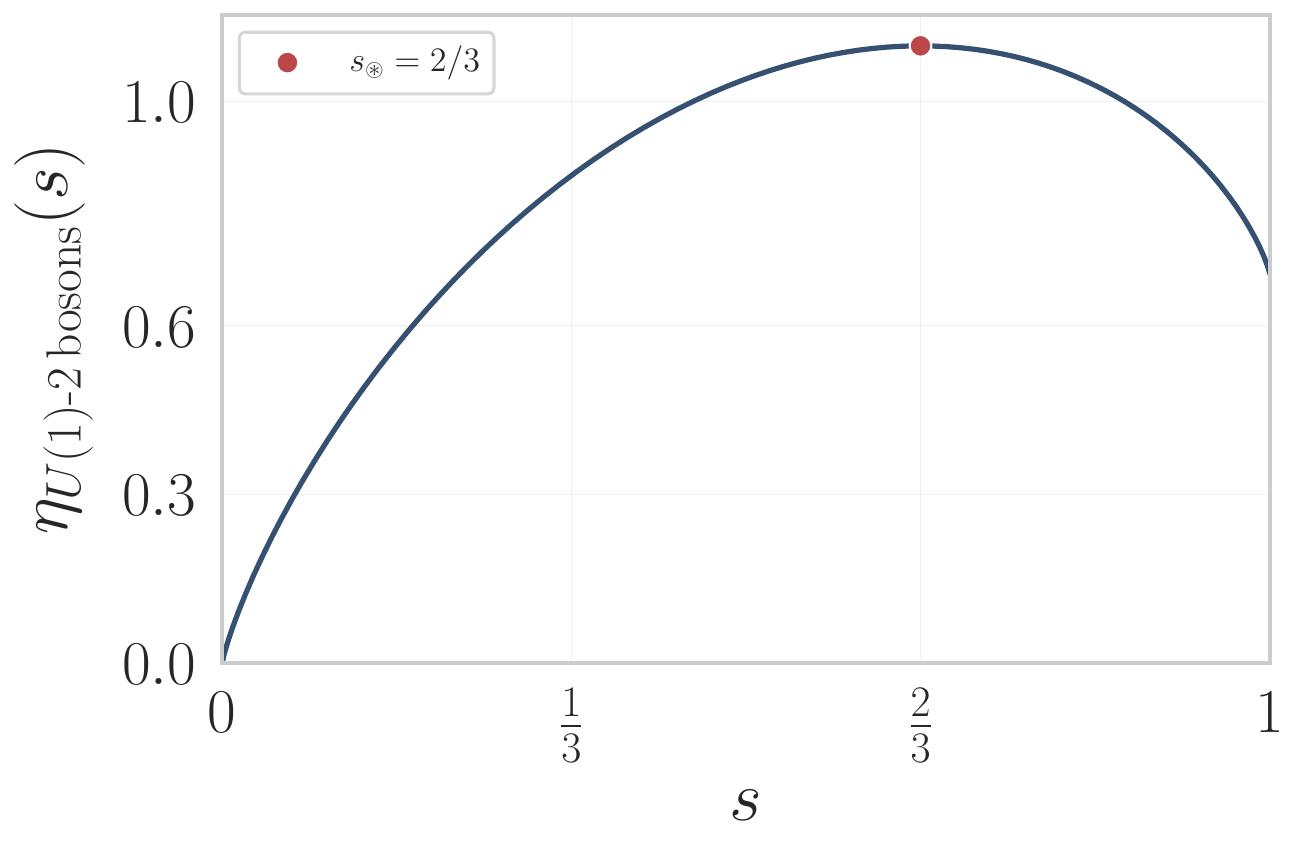}
\caption{The local entropy $\eta(s)$ for the case of:  \textnormal{(left)} $N$ qubits with fixed global $U(1)$ charge \eqref{eq:eta-U1-qubit}, \textnormal{(center)} $N$ qutrits with fixed global $U(1)$ charge \eqref{eq:eta-U1-qutrit}, and \textnormal{(right)} $N$ two-species hardcore bosons with fixed global $U(1)$ charge corresponding to fixed total number of particles \eqref{eq:eta-U1-2bosons}.}
\end{figure}

\begin{figure}[t]
\centering
\includegraphics[width=.32\textwidth]{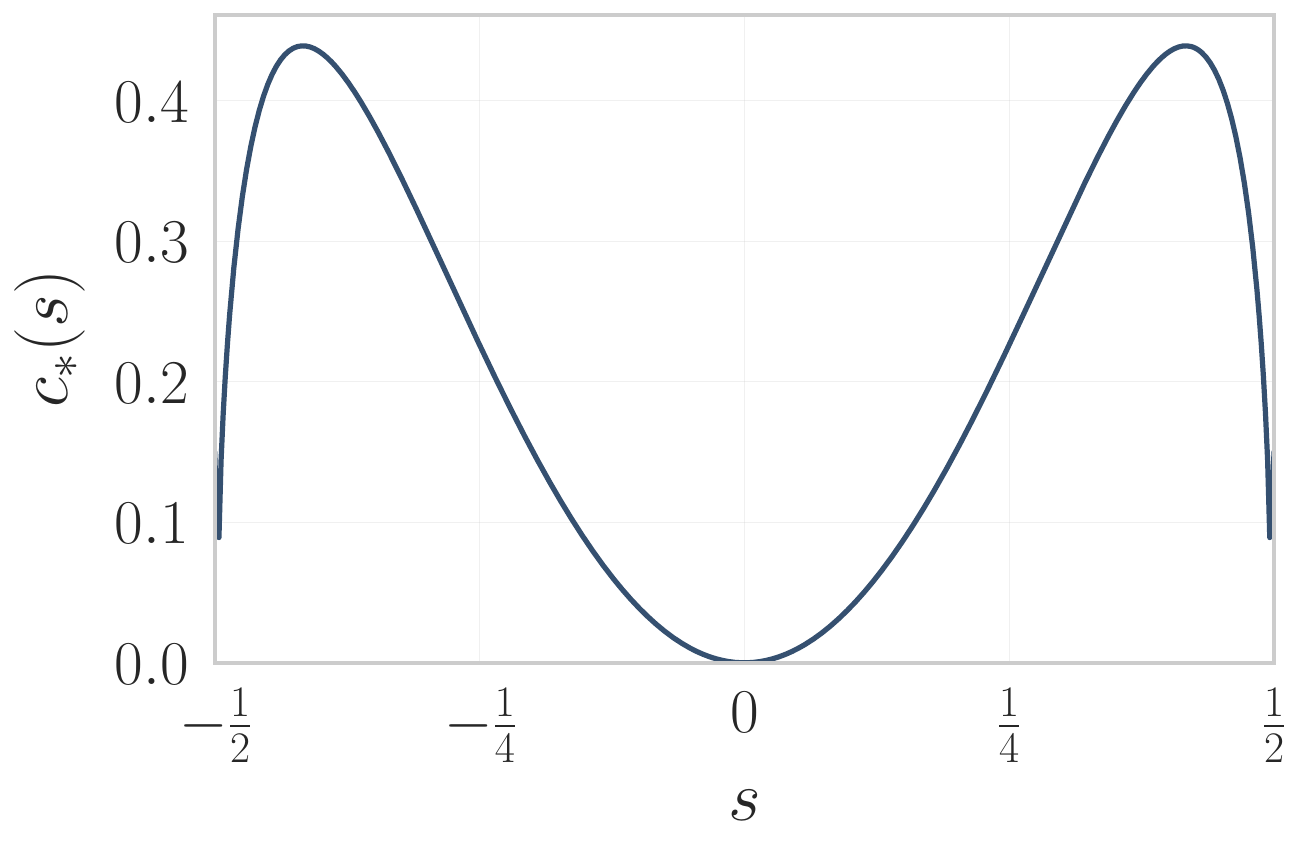}
\includegraphics[width=.32\textwidth]{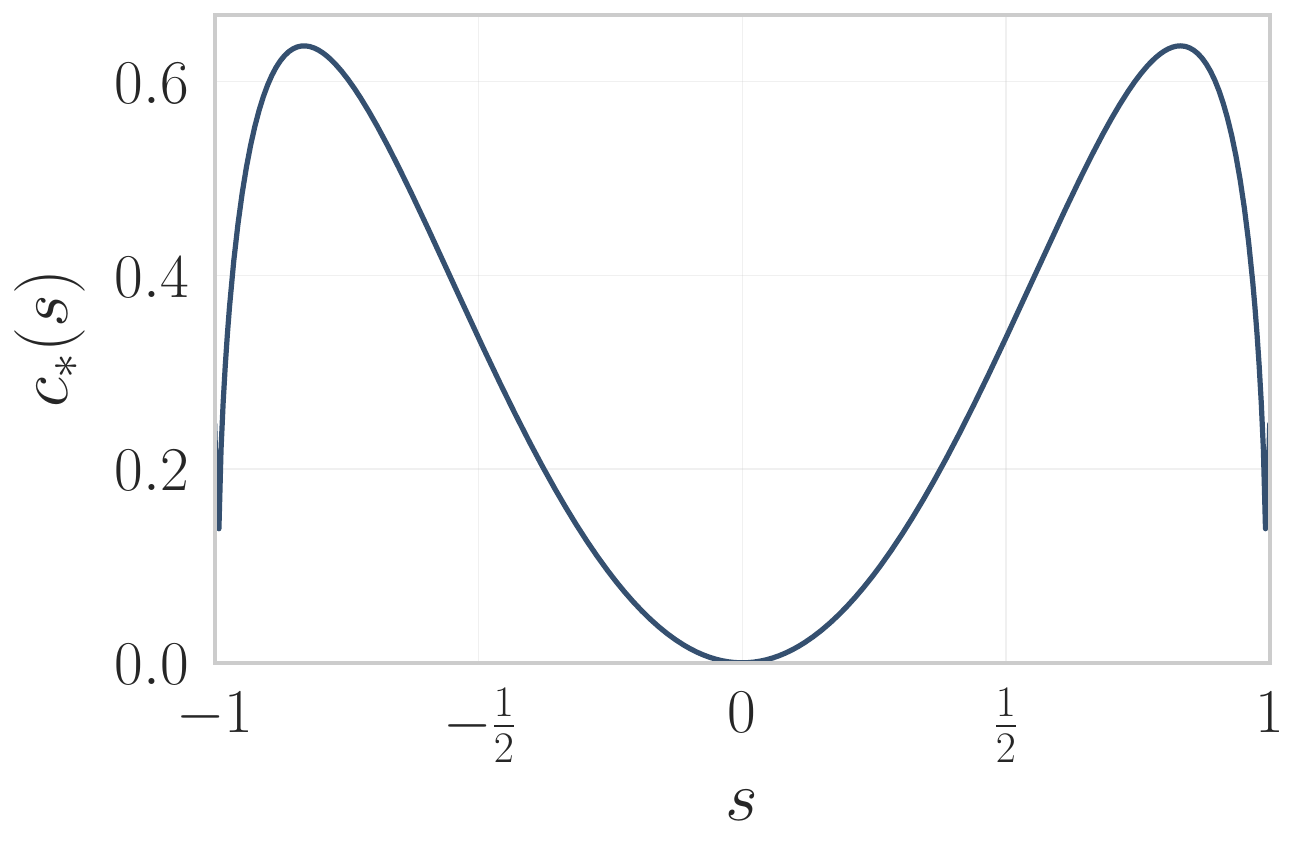}
\includegraphics[width=.32\textwidth]{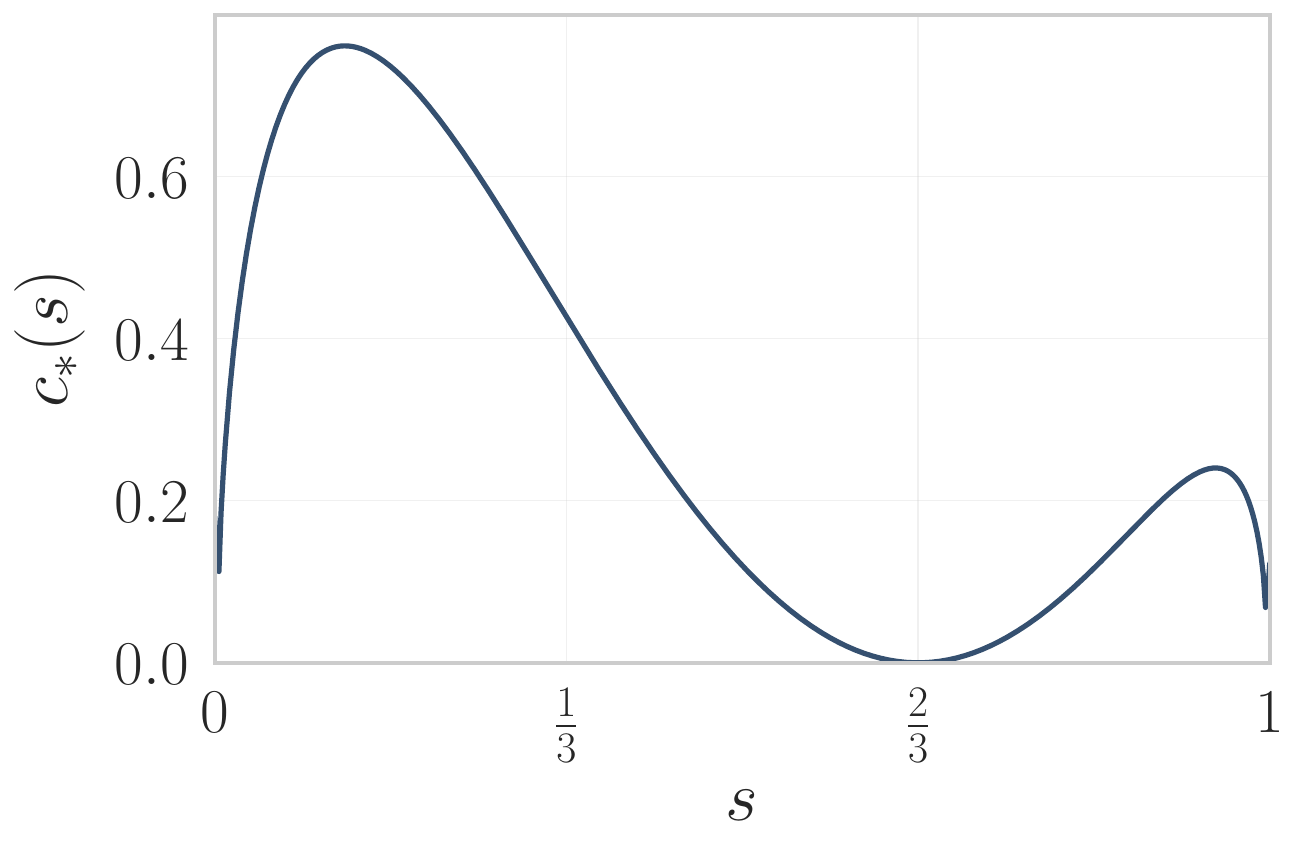}
\caption{The heat capacity $c_*(s)$ for the case of:  \textnormal{(left)} $N$ qubits with fixed global $U(1)$ charge \eqref{eq:beta-c-U1-qubit}, \textnormal{(center)} $N$ qutrits with fixed global $U(1)$ charge \eqref{eq:beta-c-U1-qutrit}, and \textnormal{(right)} $N$ two-species hardcore bosons with fixed global $U(1)$ charge corresponding to fixed total number of particles \eqref{eq:beta-c-U1-boson}.}
\end{figure}

\subsection{Example: N oscillators at fixed $U(1)$ charge (occupation number)}

The formalism discussed in the previous sections assumes that the local Hilbert space is finite-dimensional. However, as noted in  \cite{Yauk:2023wbu}, the essential requirement is weaker: the many-body Hilbert space at fixed total charge must have finite dimension. This observation allows one to treat also systems with infinite-dimensional local Hilbert spaces, provided that the fixed-charge constraint selects a finite-dimensional sector. A simple example is given by $N$ harmonic oscillators at fixed global $U(1)$ charge $Q=Ns$, where the charge is the total number of excitations. The local Hilbert space is infinite-dimensional, with $q_{\loc}=0,1,2,\ldots$, but the fixed-charge subspace has finite dimension. The saddle-point probability distribution is a geometric series,
\begin{equation}
p_{q_{\loc}}(\beta)
=
\frac{\ee^{-\beta q_{\loc}}}{\sum_{q=0}^{\infty}\ee^{-\beta q}}
=
\big(1-\ee^{-\beta}\big)\ee^{-\beta q_{\loc}}\, ,
\end{equation}
and the saddle-point condition gives
\begin{equation}
s=\sum_{q_{\loc}=0}^{\infty}q_{\loc}\,p_{q_{\loc}}(\beta_*)
=
\frac{\ee^{-\beta_*}}{1-\ee^{-\beta_*}}\, ,
\qquad
\beta_*(s)=\log\!\left(\frac{1+s}{s}\right) .
\end{equation}
Therefore, the corresponding local entropy and the associated heat capacity are
\begin{equation}
\eta_{\mathrm{osc}}(s)
=
(1+s)\log(1+s)-s\log s\, , \quad \text{and} \quad 
c_*(s)
=
s(1+s)\left(\log\frac{1+s}{s}\right)^2 ,
\end{equation}
which reproduces the result discussed in \cite{Yauk:2023wbu}.

\section{Typical entanglement entropy at fixed $SU(2)$ charge}
\label{sec:SU2}
We derive the asymptotic formulas for general systems  at fixed $SU(2)$ charge and discuss five model systems.

\subsection{Asymptotic formulas for general systems with a  $SU(2)$ charge}
Following the same steps as in the previous section, we can also compute the functions $\eta(s)$ and $\alpha_0(s)$ for the case of a conserved $SU(2)$ charge. The main difference with respect to the $U(1)$ case is that the characters of the irreducible representations of $SU(2)$ are more complicated than those of $U(1)$, and the Haar measure is also different.

The starting point is the dimension of the Hilbert space at fixed charge written using the expression \eqref{eq:dimensionDq-characters} for $SU(2)$. 
The character of the irreducible representation of charge $q=j$ is 
\begin{equation}
  \label{eq:SU2character}
  \chi^{(j)}(g)=\sum_{m=-j}^{j} \ee^{\ii m \theta}=\frac{\sin((2j+1)\theta/2)}{\sin(\theta/2)} \ ,
\end{equation}
where $\theta$ parametrizes the diagonal group elements $g=\ee^{\ii \theta \sigma_z/2}$. The character is invariant under group conjugation, so we can restrict to the diagonal elements without any loss of generality. In terms of the angular variable $\theta\in [0,2\pi)$, the Haar measure for $SU(2)$ is given by $d\mu(g)=\frac{1}{\pi}\sin^2(\theta/2) d\theta$, which is the measure on the maximal torus of $SU(2)$, normalized such that $\int_0^{2\pi} d\mu(g)=1$. The character of the local Hilbert-space representation is
\begin{equation}
\chi_\loc(g)=\sum_{j_{\loc}} a_{j_{\loc}}   \chi^{(j_\loc)}(g)\,=\sum_{j_{\loc}\,m_{\loc}}a_{j_{\loc}}  \,\ee^{\ii m_{\loc} \theta}\,,
\end{equation}
where, in this compact notation, the magnetic quantum number $m_{\loc}$ runs from $-j_{\loc}$ to $+j_{\loc}$ in integer steps which could be evaluated explicitly as in \eqref{eq:SU2character}, but it is more convenient to leave in this form for the moment. With these definitions, the dimension of the Hilbert space at fixed charge $j=Ns$ can be written as
\begin{equation}
\label{eq:general-SU2-dim}
D_{Ns}=
\int_0^{2\pi}  
\frac{\sin((2Ns+1)\theta/2)}{\sin(\theta/2)} 
\Big(\sum_{j_{\loc}\,m_{\loc}} a_{j_{\loc}} \,\ee^{\ii m_{\loc} \theta}\Big)^N \,\frac{1}{\pi}\sin^2(\theta/2) \,d\theta\, .
\end{equation}
To evaluate the integral in the large $N$ limit using the Laplace method, we can manipulate the integrand to write it in a more suitable form. First, we can split $\sin((2Ns+1)\theta/2)$ into two exponentials and write the integrand as a sum of two terms. Then, we can perform the change of variable $\theta \to -\theta$ in one of the two terms, which leaves the integration measure and the local character invariant. This allows us to write the dimension as
\begin{equation}
D_{Ns}=
\int_0^{2\pi} \frac{d\theta}{2\pi} \big(1 -\ee^{-\ii \theta}\big) 
\,\ee^{-\ii N s \theta}
\Big(\sum_{j_{\loc}\,m_{\loc}} a_{j_{\loc}} \,\ee^{\ii m_{\loc} \theta}\Big)^N 
=
\int_0^{2\pi} \frac{d\theta}{2\pi} \big(1 -\ee^{-\ii \theta}\big) 
\,\ee^{N \tilde\eta(-\ii \theta,\,s)} \, .
\end{equation}
Here we have defined the function
\begin{equation}
\tilde\eta(\beta,s)=\log\Big(\sum_{j_{\loc}\,m_{\loc}} a_{j_{\loc}} \,\ee^{-m_{\loc} \beta}\Big) + s \beta\,.
\end{equation}
The integral is dominated by the stationary point $\beta_*(s)$, determined by the condition 
\begin{equation}
\label{eq:saddleSU2}
\partial_\beta \tilde\eta(\beta,s)\vert_{\beta_*(s)}=0 
\quad \rightarrow \quad 
\frac{\sum_{j_{\loc} m_{\loc}} m_{\loc} a_{j_{\loc}} \,\ee^{-m_{\loc} \beta_*(s)}}{\sum_{j_{\loc} m_{\loc}} a_{j_{\loc}} \,\ee^{-m_{\loc} \beta_*(s)}} = s \,.
\end{equation}
Formally, this is the same condition as in the $U(1)$ case, but now the sum runs over the magnetic quantum numbers $m_{\loc}$ of the irreducible representations of $SU(2)$, which are weighted by their multiplicity $a_{j_{\loc}}$. Also in this case, the stationary point $\beta_*(s)$ does not lie on the original integration contour, which is the imaginary axis, but it can be reached by deforming the contour in the complex plane.

A few comments are in order about the stationary-point condition \eqref{eq:saddleSU2}. Positivity of the charge implies that the inverse temperature $\beta_*(s)$ is negative, which in turn implies that the function $\eta(s)$ is monotonically decreasing (its derivative is negative). Since $\eta(s)$ is also concave (its second derivative is negative), it follows that the maximum of $\eta(s)$ is attained at $s_\circledast=0$. This extreme point is special, since it corresponds to the case of zero charge. The leading order of the dimension $D_{Ns}$ at $s=s_\circledast$ vanishes and the next-to-leading order becomes relevant. We do not report the explicit form of the entanglement entropy at $s=s_\circledast$ here, since it is not relevant for the present calculations. We refer to Sec.~6.2 of \cite{Bianchi:2024aim} for an analysis of these extremal cases.

Evaluating the function $\tilde\eta(\beta,s)$ at the stationary point $\beta_*(s)$ gives the function $\eta(s)$ appearing in \eqref{eq:dimensionDq-Asympt}. Completing the stationary-point calculation of $D_{Ns}$, and simplifying the result using the properties of $\eta(s)$, we find
\begin{equation}
D_{Ns} = \big(1 -\ee^{\eta'(s)}\big)  \sqrt{\frac{-\eta''(s)}{2\pi N}} \big(1+\O(1/N)\big)\,\ee^{N \eta(s)}\,.
\end{equation}
From this expression for $D_{Ns}$ we can read the functions $\eta(s)$ and $\alpha_0(s)$ defined in \eqref{eq:dimensionDq-Asympt}:
\begin{equation}
\eta(s) = \log\Big(\sum_{j_{\loc}\, m_{\loc}} a_{j_{\loc}} \,\ee^{-m_{\loc} \beta_*(s)}\Big) + s \beta_*(s)\, , \quad \text{and}\quad \alpha_0(s) = 1 -\ee^{\eta'(s)} \,.
\end{equation}
The next-to-leading order correction $\alpha_1(s)$ can be computed in a similar way. We do not report its value as it cancels out in the expression of the average entanglement entropy at order $\O(N^0)$.

Let us summarize the generic features of the case with fixed global $SU(2)$ charge. The probability distribution over the magnetic quantum number $m_{\loc}$ in the irreducible representation
$j_{\loc}$ is
\begin{equation}
p_{j_{\loc},m_{\loc}}(\beta)
=\frac{a_{j_{\loc}} \,\ee^{-\beta m_{\loc}}}{\displaystyle\sum_{j_{\loc}\,m_{\loc}}
a_{j_{\loc}} \,\ee^{-\beta m_{\loc}}}\,.
\end{equation}
The inverse temperature $\beta_*(s)$ is fixed by imposing that the average magnetic quantum number equals the charge density $s$,
\begin{equation}
\label{eq:beta-star-SU2}
\sum_{j_{\loc}\,m_{\loc}} m_{\loc}\;p_{j_{\loc},m_{\loc}}(\beta_*)\;=\;s\,.
\end{equation}
The function $\eta(s)$ is $\log(k)$ minus the relative entropy of the probability distribution at temperature $\beta_*(s)^{-1}$ with respect to its infinite temperature limit $p_{j_{\loc},m_{\loc}}(0)=a_{j_{\loc}}/k$, where
$k=\sum_{j_{\loc}}(2j_{\loc}+1)a_{j_{\loc}}$ is the local Hilbert space dimension:
\begin{equation}
\eta(s)=\log(k)-\sum_{j_{\loc}\,m_{\loc}}p_{j_{\loc},m_{\loc}}(\beta_*(s))
\log\!\Big(\frac{p_{j_{\loc},m_{\loc}}(\beta_*(s))}{p_{j_{\loc},m_{\loc}}(0)}\Big)\,.
\end{equation}
In terms of $\eta$, the inverse temperature is the first derivative,
\begin{equation}
\beta_*(s)=\eta'(s)\,,
\end{equation}
motivating the interpretation of $\eta(s)$ as a thermodynamic potential, and the heat capacity is
\begin{equation}
c_*(s)=\frac{\eta'(s)^2}{-\eta''(s)}\,.
\end{equation}
The key difference with respect to the $U(1)$ case is that the coefficient $\alpha_0(s)$ is no longer trivial. It takes the $SU(2)$-specific value
\begin{equation}
\label{eq:alpha0-SU2}
\alpha_0(s)=1-\ee^{\eta'(s)}=1-\ee^{\beta_*(s)}\,.
\end{equation}
Since $s\geq 0$ implies $\beta_*(s)\leq 0$, the prefactor satisfies $0\leq\alpha_0(s)<1$ throughout the physical domain, with $\alpha_0(s)=0$ only at the boundary $s=0$ where the stationary-point formula breaks down (See Sec.~6.2 of \cite{Bianchi:2024aim} for the analysis of this extremal case). The condition $\eta'(s_\circledast)=0$ is therefore only met at $s_\circledast=0$ and, as a result, the $\delta_{s,s_\circledast}$ contribution present in the $U(1)$ formula at $f=\frac{1}{2}$ does not appear for non-extremal $SU(2)$ charge. Substituting $\alpha_0(s)=1-\ee^{\beta_*(s)}$ into \eqref{eq:entropyasymptotic}, we find that the average entanglement entropy at fixed global $SU(2)$ charge $j=Ns$ takes the form
\begin{align}
\label{eq:entropyasymptotic-SU2}
\setlength{\fboxsep}{9pt}
\boxed{
\!\!\langle S_A\rangle_{Ns} \!=\!
\begin{cases}
\;\textstyle
  f \eta(s) N + \frac{\log(1-f)+f}{2}
  + \log\!\big(1-\ee^{\beta_*(s)}\big)
  + (1-f)\frac{\beta_*(s)\,\ee^{\beta_*(s)}}{1-\ee^{\beta_*(s)}}
  + \O(N^{-1})\!\!
& f<\tfrac{1}{2}\\[1.2em]
\;\textstyle
  \tfrac{1}{2}\eta(s) N
  -\sqrt{\frac{c_*(s)}{2\pi}}\sqrt{N}
  +\frac{\log(1/2)+1/2}{2}
  +\tfrac{1}{2}\log\!\big(1-\ee^{\beta_*(s)}\big)
  + \O(N^{-1/2})
& f=\tfrac{1}{2}\\[1.2em]
\;\textstyle
  (1-f)\eta(s) N + \frac{\log(f)+1-f}{2}
  - (1-f)\frac{\beta_*(s)\,\ee^{\beta_*(s)}}{1-\ee^{\beta_*(s)}}
  + \O(N^{-1})
& f>\tfrac{1}{2}
\end{cases}
}
\end{align}
The variance of the entanglement entropy with a fixed global $SU(2)$ charge is
\begin{equation}
\label{eq:variance-SU2}
(\Delta S_A)^2_{Ns}=\left(\Big(\sqrt{2\pi}\,f(1-f)-\tfrac{1}{\sqrt{2\pi}}\delta_{f,\frac{1}{2}}\Big)\frac{c_*(s)^{3/2}}{(1-\ee^{\beta_*(s)})\,|\beta_*(s)|}\,N^{3/2}+\O(N)\right)\ee^{-N\eta(s)}\, ,
\end{equation}
which is exponentially suppressed in $N$ for any charge density $s$ such that $\eta(s)\neq 0$. Therefore, the average entanglement entropy $\langle S_A\rangle_{Ns}$ is typical in the thermodynamic limit.

In the rest of this section we provide various examples corresponding to different structures of the local Hilbert space $\mathcal{H}_\loc$, for which we provide explicit expressions of the local entropy $\eta(s)$.

\begin{figure}[t]
\centering
\includegraphics[width=.32\textwidth]{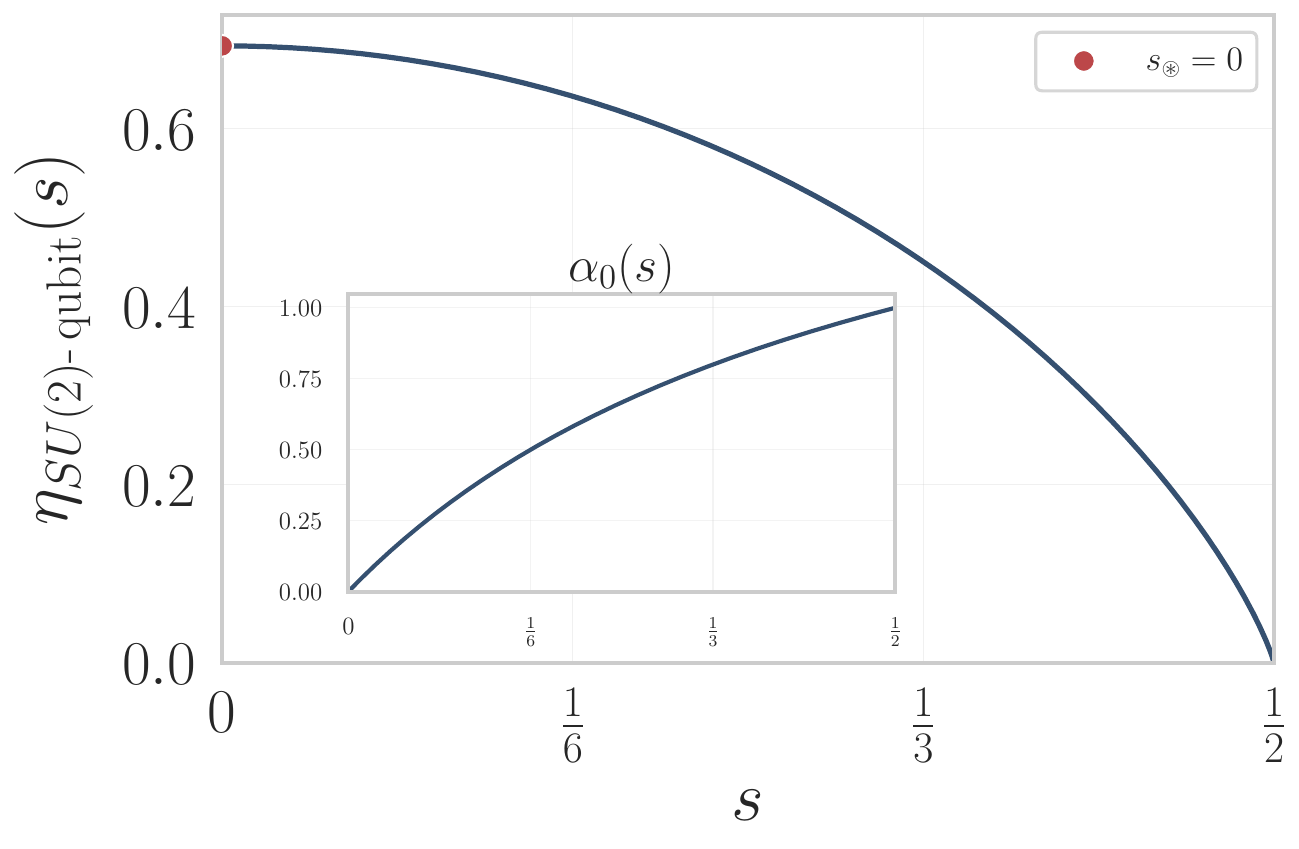}
\includegraphics[width=.32\textwidth]{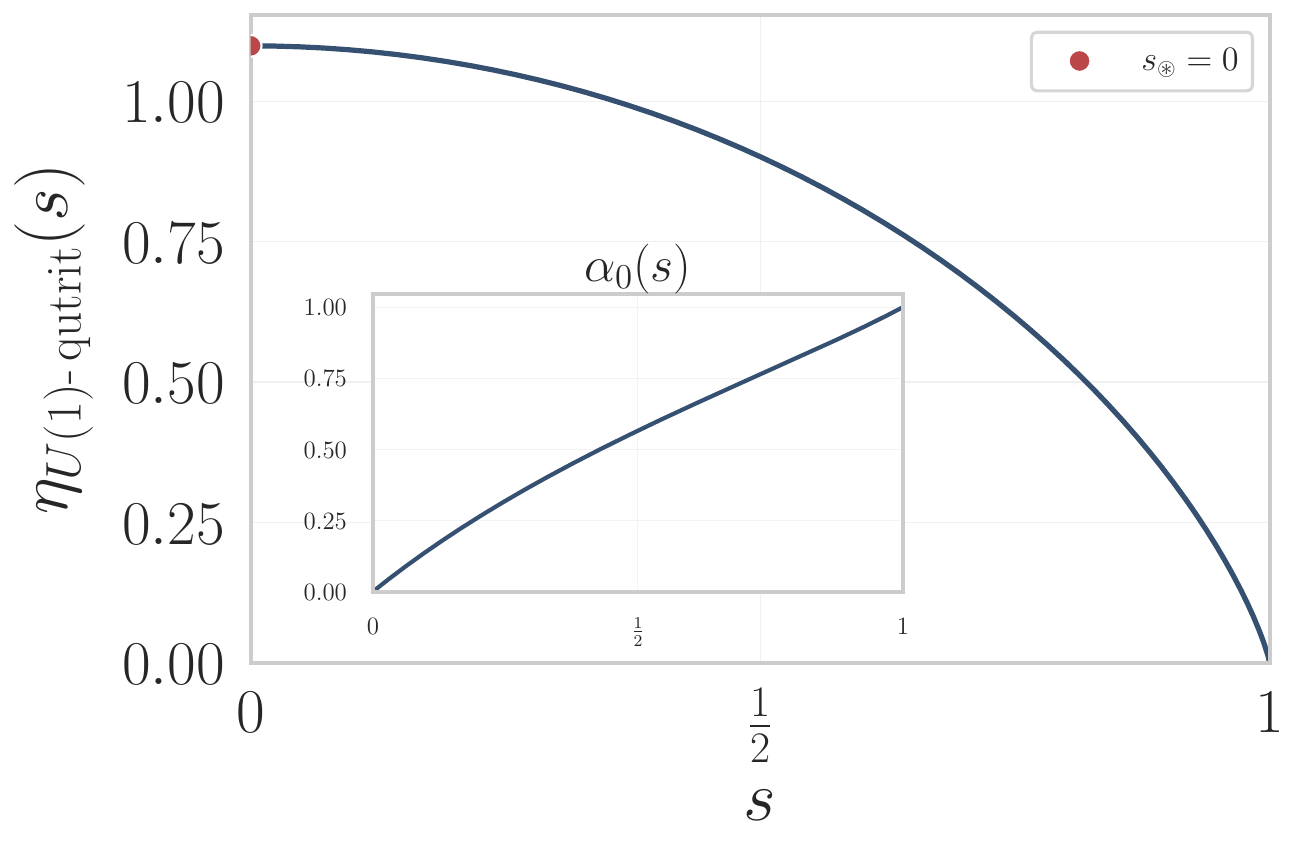}
\includegraphics[width=.32\textwidth]{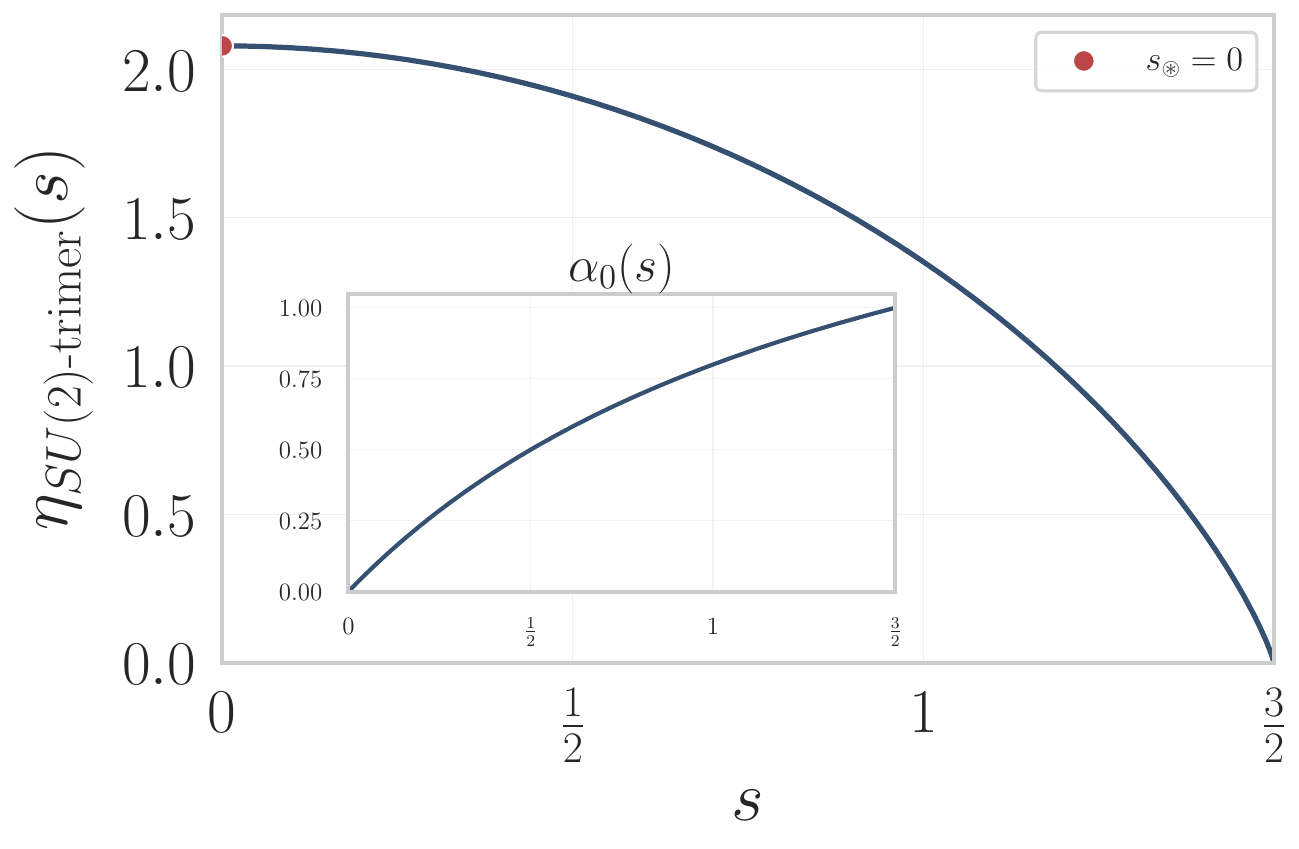}
\caption{The function $\eta(s)$ for the case of $N$ qubits with $SU(2)$ charge \eqref{eq:eta-SU2-qubit} (left), for the case of $N$ spin-$1$ (qutrits) with $SU(2)$ charge \eqref{eq:eta-SU2-qutrit} (center), and for the case of $N$ spin-$\frac{1}{2}$ trimers with $SU(2)$ charge \eqref{eq:eta-SU2-trimer} (right).}
\end{figure}

\subsection{Example: $N$ qubits at fixed global $SU(2)$ charge (spin $1/2$)}
\label{sec:qubit-SU2}
We consider the case of $N$ qubits. Each qubit Hilbert space carries the local $SU(2)$ charge $j_{\loc}=1/2$. The local Hilbert space decomposes as 
\begin{equation}
\mathcal{H}_{\loc}=\mathcal{H}^{(1/2)}_{\mathrm{irrep}}\,,
\end{equation}
with trivial multiplicity. The local Hilbert space is two-dimensional and a basis of states is labeled by the magnetic quantum numbers $m_{\loc}=\pm 1/2$. We compute the average entanglement entropy of this $N$-body system at fixed global $SU(2)$ charge $j=Ns$ and fixed subsystem fraction $f$. The spectrum of the physical Hamiltonian of a quantum chaotic spin-$1/2$ Heisenberg chain decomposes into sectors of fixed spin $j$ and has the same structure as the one discussed here 
\cite{Bianchi:2024aim,Patil:2023wdw,Patil:2025ump,Chakraborty:2025ziy,Wu:2026jxe,Yauk:2026quy}.

The probability distribution entering the definition of $\eta(s)$ is
\begin{equation}
  p_{j_{\loc},m_{\loc}}(\beta)=\frac{\ee^{-\beta m_{\loc}}}{\ee^{-\beta/2}+\ee^{\beta/2}}\,.
\end{equation}
The inverse temperature $\beta$ is fixed by the stationary-point condition \eqref{eq:beta-star}, which gives
\begin{equation}
  \frac{\tfrac{1}{2}\ee^{-\beta_*(s)/2}-\tfrac{1}{2}\ee^{\beta_*(s)/2}}{\ee^{-\beta_*(s)/2}+\ee^{\beta_*(s)/2}}= 
  s\,, \quad \rightarrow \quad \beta_*(s) = -2\mathrm{arctanh}(2s)\,,
\end{equation}
with $0<s<1/2$. We can then write the probabilities $p_{j_{\loc},m_{\loc}}(\beta_*(s))$ in terms of $s$ directly
\begin{equation*}
p_{1/2,-1/2}(s)=\frac{1-2s}{2}\,,\qquad p_{1/2,+1/2}(s)=\frac{1+2s}{2}\,.
\end{equation*}
With these definitions, the local entropy $\eta(s)$ can be computed explicitly and takes the form
\begin{equation}
\eta_{{SU(2)-\mathrm{qubit}}}(s)\,=\,-\frac{1-2s}{2}\log\Big(\frac{1-2s}{2}\Big)-\frac{1+2s}{2}\log\Big(\frac{1+2s}{2}\Big)\,.
\label{eq:eta-SU2-qubit}
\end{equation}
This is the same function $\eta(s)$ found for the $U(1)$ qubit \eqref{eq:eta-U1-qubit}, while the group-dependent prefactor is different and is given by
\begin{equation}
\alpha_0(s)=1-\ee^{\eta'(s)}=\frac{4s}{1+2s}\,.
\end{equation}
The average entanglement entropy at fixed global spin can then be obtained by plugging \eqref{eq:eta-SU2-qubit} and $\alpha_0(s)$ into the general formulas \eqref{eq:entropyasymptotic}. The result reproduces the known expression for the typical entanglement entropy of $N$ qubits at fixed total spin $j=Ns$ found in \cite{Bianchi:2024aim,Patil:2023wdw,Patil:2025ump,Chakraborty:2025ziy,Wu:2026jxe,Yauk:2026quy}.

From $\eta(s)$ we can compute the temperature and the local heat capacity:
\begin{equation}
\label{eq:beta-c-SU2-qubit}
\beta_*(s)=\log\Big(\frac{1 - 2 s}{1 + 2 s}\Big) \, , \qquad
c_*(s)=\big(\tfrac{1}{4}-s^2\big)\, \Big(\log\frac{1 - 2 s}{1 + 2 s}\Big)^2 \, .
\end{equation}

\subsection{Example: $N$ qutrits at fixed global $SU(2)$ charge (spin $1$)}
\label{sec:qutrit-SU2}
We consider the case of $N$ qutrits, $k=3$. Each qutrit Hilbert space carries the local $SU(2)$ charge $j_{\loc}=1$. The local Hilbert space decomposes as 
\begin{equation}
\mathcal{H}_{\loc}=\mathcal{H}^{(1)}_{\mathrm{irrep}}
\end{equation}
with trivial multiplicity $a_1=1$.  The local Hilbert space is three-dimensional and a basis of states is labeled by the magnetic quantum numbers $m_{\loc}=-1,0,+1$. We compute the average entanglement entropy of this $N$-body system at fixed global $SU(2)$ charge $j=Ns$ and fixed subsystem fraction $f$. The spectrum of the physical Hamiltonian of a quantum chaotic spin-$1$ Heisenberg chain \cite{Patil:2023wdw} decomposes into sectors of fixed spin $j$  and has the same structure as the one discussed here.

The probability distribution entering the definition of $\eta(s)$ is
\begin{equation}
  p_{1,m_{\loc}}(\beta)=\frac{\ee^{-\beta m_{\loc}}}{\ee^{\beta}+1+\ee^{-\beta}}\,.
\end{equation}
The inverse temperature $\beta$ is fixed by the saddle-point condition \eqref{eq:saddleSU2},
which gives
\begin{equation}
  \frac{\ee^{-\beta_*(s)}-\ee^{\beta_*(s)}}{\ee^{-\beta_*(s)}+1+\ee^{\beta_*(s)}}
  =s\, , 
  \quad\rightarrow\quad
  \beta_*(s)=\log\!\Big(\frac{-s+\sqrt{4-3s^2}}{2(1+s)}\Big)\,,
\end{equation}
with the charge density $0<s<1$. With these definitions, the probabilities $p_{1,m_{\loc}}(\beta_*(s))$ can be written explicitly as
\begin{equation*}
p_{1,-1}(s)=\frac{4-3s-\sqrt{4-3s^2}}{6}\,,\ 
p_{1,0}(s)=\frac{-1+\sqrt{4-3s^2}}{3}\,,\ 
p_{1,+1}(s)=\frac{4+3s-\sqrt{4-3s^2}}{6}\,.
\end{equation*}
The function $\eta(s)=-\sum_{m_{\loc}}p_{1,m_{\loc}}\log p_{1,m_{\loc}}$ can be computed explicitly and takes the form
\begin{equation}
\label{eq:eta-SU2-qutrit}
\begin{aligned}
\eta_{{SU(2)\text{-qutrit}}}(s)\,=\,
&-\frac{4-3s-\sqrt{4-3s^2}}{6}\log\!\Big(\frac{4-3s-\sqrt{4-3s^2}}{6}\Big)\\
&-\frac{-1+\sqrt{4-3s^2}}{3}\log\!\Big(\frac{-1+\sqrt{4-3s^2}}{3}\Big)\\
&-\frac{4+3s-\sqrt{4-3s^2}}{6}\log\!\Big(\frac{4+3s-\sqrt{4-3s^2}}{6}\Big)\,.
\end{aligned}
\end{equation}
This is the same function $\eta(s)$ as for the $U(1)$ qutrit \eqref{eq:eta-U1-qutrit}, since the local charge spectrum $m_{\loc}\in\{-1,0,+1\}$ and the multiplicities coincide in the two cases. The group-dependent prefactor, however, is different and is given by 
\begin{equation}
\alpha_0(s)=1-\ee^{\eta'(s)}=1-\ee^{\beta_*(s)}=\frac{2+3s-\sqrt{4-3s^2}}{2(1+s)}\,.
\end{equation}
The entanglement entropy at fixed total spin can then be obtained by plugging \eqref{eq:eta-SU2-qutrit} and $\alpha_0(s)$ into the general formulas
\eqref{eq:entropyasymptotic}.

From $\eta(s)$ we can compute the temperature and the local heat capacity:
\begin{align}
\beta_*(s)&=\log\Big(\frac{-s + \sqrt{4 - 3 s^2}}{2 (1 + s)}\Big)\, , \\[.5em] 
c_*(s)&= \frac{\sqrt{4-3 s^2}\big(\sqrt{4-3 s^2}-1\big)}{12} \Big(\log\frac{4+3 s-\sqrt{4-3 s^2}}{4-3s-\sqrt{4-3 s^2}}\,\Big)^2 \, .
\label{eq:beta-c-SU2-qutrit}
\end{align}

\subsection{Example: $N$ trimers of spin-$\frac{1}{2}$ at fixed $SU(2)$ charge}
We consider the case of $N$ qukits, $k=8$. Each local Hilbert space is made of three spin-$1/2$ degrees of freedom transforming under the diagonal action of $SU(2)$. In this case the local Hilbert space decomposes as
\begin{equation}
\mathcal H_{\loc}
=
\mathcal{H}_{\mathrm{irrep}}^{(1/2)}\otimes \mathcal{H}_{\mathrm{irrep}}^{(1/2)}\otimes \mathcal{H}_{\mathrm{irrep}}^{(1/2)}
=
\mathcal{H}_{\mathrm{irrep}}^{(1/2)}\oplus \mathcal{H}_{\mathrm{irrep}}^{(1/2)}\oplus \mathcal{H}_{\mathrm{irrep}}^{(3/2)}\,.
\end{equation}
Equivalently, the only non-vanishing multiplicities are $a_{1/2}=2$, $a_{3/2}=1$. We compute the average entanglement entropy of this $N$-body system at fixed global $SU(2)$ charge $j=Ns$ and fixed subsystem fraction $f$. Microscopically, each of the bodies can be viewed as
a local trimer of three exchange-coupled spin-$1/2$ moments, as realized for instance in triangular molecular magnets~\cite{Cage:2003} and in semiconductor double-quantum-dot
spin trimers~\cite{Jang:2025}.

The probability distribution entering the asymptotic analysis is
\begin{equation}
p_{j_{\loc},m_{\loc}}(\beta)=\frac{a_{j_{\loc}}\,\ee^{-\beta m_{\loc}}}{\ee^{-3\beta/2}+3\ee^{-\beta/2}+3\ee^{\beta/2}+\ee^{3\beta/2}}\,.
\end{equation}
The inverse temperature $\beta_*(s)$ is fixed by the saddle-point condition \eqref{eq:beta-star}, which gives
\begin{equation}
\frac{
\frac{3}{2}\ee^{-3\beta_*(s)/2}
+\frac{3}{2}\ee^{-\beta_*(s)/2}
-\frac{3}{2}\ee^{\beta_*(s)/2}
-\frac{3}{2}\ee^{3\beta_*(s)/2}
}{
\ee^{-3\beta_*(s)/2}
+3\ee^{-\beta_*(s)/2}
+3\ee^{\beta_*(s)/2}
+\ee^{3\beta_*(s)/2}
}
=
-\frac{3}{2}\tanh\!\big(\beta_*(s)/2\big)
=
s\,,
\end{equation}
so that
\begin{equation}
\beta_*(s)=-2\,\mathrm{arctanh}\!\left(\tfrac{2s}{3}\right)\,,
\qquad
0<s<\tfrac{3}{2}\,.
\end{equation}
With these definitions, the probabilities $p_{j_{\loc},m_{\loc}}(\beta_*(s))$ can be written explicitly in terms of $s$ as
\begin{equation*}
  \begin{aligned}
    &p_{3/2,-3/2}(s)=\frac{(3-2s)^3}{6^3}\,,
    &&p_{3/2,-1/2}(s)=\frac{(3-2s)^2(3+2s)}{6^3}\,,\\
    &p_{3/2,+1/2}(s)=\frac{(3-2s)(3+2s)^2}{6^3}\,,
    &&p_{3/2,+3/2}(s)=\frac{(3+2s)^3}{6^3}\,,\\
    &p_{1/2,-1/2}(s)=2\frac{(3-2s)^2(3+2s)}{6^3}\,,
    &&p_{1/2,+1/2}(s)=2\frac{(3-2s)(3+2s)^2}{6^3}\,.
  \end{aligned}
\end{equation*}
The function $\eta(s)$ can be computed explicitly as 
\begin{equation}
  \eta(s)= -\sum_{m_\loc=-3/2}^{3/2} p_{3/2,m_\loc}(s) \log p_{3/2,m_\loc}(s) - \sum_{m_\loc=-1/2}^{1/2} p_{1/2,m_\loc}(s) \log \frac{p_{1/2,m_\loc}(s)}{2} \, ,
\end{equation}
and takes the explicit form
\begin{equation}
\eta_{SU(2)\text{-trimer}}(s)\,=\,-\frac{3-2s}{2}\log\Big(\frac{3-2s}{6}\Big)-\frac{3+2s}{2}\log\Big(\frac{3+2s}{6}\Big)\,.
\label{eq:eta-SU2-trimer}
\end{equation}
Equivalently, this can be written as
\begin{equation}
\eta_{SU(2)\text{-trimer}}(s)=3\,\eta_{{SU(2)-\mathrm{qubit}}}(s/3)\, ,
\end{equation}
showing that this example is equivalent to three independent spin-$1/2$ grouped into a single local $SU(2)$ representation. The group-dependent prefactor is
\begin{equation}
\alpha_0(s)=1-\ee^{\eta'(s)}=1-\ee^{\beta_*(s)}=\frac{4s}{3+2s}\,.
\end{equation}
The entanglement entropy at fixed total spin can then be obtained by plugging \eqref{eq:eta-SU2-trimer} and $\alpha_0(s)$ into the general formulas \eqref{eq:entropyasymptotic}.

From $\eta(s)$ we can compute the temperature and the local heat capacity:
\begin{equation}
\beta_*(s)= \log\Big(\frac{3 - 2 s}{3 + 2 s}\Big) \, , \qquad
c_*(s)= \Big(\frac{3}{4}-\frac{s^2}{3}\Big)\Big(\!\log\frac{3 - 2 s}{3 + 2 s}\,\Big)^2 \, .
\label{eq:beta-c-SU2-trimer}
\end{equation}

\subsection{Example: $N$ $n$-mers of spin-$\frac{1}{2}$ at fixed $SU(2)$ charge}

The same computation as in the previous section extends immediately to a local $n$-mer made of $n$ spin-$1/2$ degrees of freedom. In this case the saddle-point condition gives
\begin{equation}
s=-\frac n2\tanh\!\big(\beta_*(s)/2\big) \, , \qquad 
\beta_*(s)
=
-2\,\mathrm{arctanh}\!\left(\frac{2s}{n}\right)
=
\log\!\left(\frac{n-2s}{n+2s}\right) \, .
\end{equation}
The local probabilities are the same as those of $n$ independent spin-$1/2$ variables with mean total magnetic charge $s$
\begin{equation} 
  p^{(n)}_{j_{\loc},m_{\loc}}(s) = a^{(n)}_{j_{\loc}}\; p_{1/2, + 1/2}(s)^{\frac n2+m_{\loc}}\; p_{1/2,- 1/2}(s)^{\frac n2-m_{\loc}}\,,  
\end{equation}
where $p_{1/2,\pm 1/2}(s/n)=\frac{n\pm 2s}{2n}$ (computed in Sec.~\ref{sec:qubit-SU2}) and $a^{(n)}_{j_{\loc}}$ is the multiplicity of the spin-$j_{\loc}$ representation in the decomposition of $n$ spin-$1/2$ representations. Therefore the entropy function is
\begin{equation}
\eta_{SU(2)\text{-$n$-mer}}(s)
=
n\,\eta_{SU(2)\text{-}\mathrm{qubit}}(s/n)\, ,
\end{equation}
namely
\begin{equation}
\eta_{SU(2)\text{-$n$-mer}}(s)
=
-\Big(\frac n2-s\Big)
\log\!\Big(\frac{n-2s}{2n}\Big)
-\Big(\frac n2+s\Big)
\log\!\Big(\frac{n+2s}{2n}\Big)\, .
\end{equation}

This result illustrates that the local entropy $\eta(s)$ cannot discriminate, by itself, whether $Q$ is  the charge of  an Abelian symmetry or a non-Abelian one. It only depends on the spectrum of the charge and on its degeneracy, or equivalently on the dimensions and multiplicities appearing in the local representation. Consequently, the same function $\eta(s)$ for a non-Abelian $SU(2)$ model can be reproduced by an Abelian $U(1)$ model which has the same local charge sectors and degeneracies. For example, the spin-$j$ representation of $SU(2)$ has charge sectors $m=-j, \cdots ,+j$ with unit degeneracy, and is therefore indistinguishable at the level of $\eta(s)$ from a $(2j+1)$-level $U(1)$ system with charges $q=-j, \cdots ,+j$. 

As a result, the leading order in the entanglement entropy is the same for the Abelian and the non-Abelian case and it is given by the same extensive contribution  $fN\eta(s)$ for $f<1/2$. On the other hand, at the order $\O(N^0)$ there is both a universal term  $(\log(1-f) +f)/2$ and a group-dependent term. Specifically, the group dependent term in \eqref{eq:entropyasymptotic-SU2} is given by
\begin{equation}
\langle S_A\rangle_{Ns}^{SU(2)}-\langle S_A\rangle_{Ns}^{U(1)}= \log\!\big(1-\ee^{\eta'(s)}\big)
  + (1-f)\frac{\eta'(s)\,\ee^{\eta'(s)}}{1-\ee^{\eta'(s)}} +\O(N^{-1})\quad\,\quad f<1/2\,.
\end{equation}
Therefore, as $\eta'(s)<0$ for $s>0$, the non-abelian model has smaller entropy than the abelian one for $f<1/2$. On the other hand, for $f>1/2$, the non-abelian model has larger entropy than the abelian one. It would be interesting to investigate the consequencies of this observation for the results of
\cite{Majidy:2022kzx}.

\subsection{Example: $N$ mixed spin-$\frac{1}{2}$--spin-$1$ dimers at fixed $SU(2)$ charge}
\label{sec:mixed-spin-SU2}

We now consider a six-dimensional local Hilbert space carrying the tensor-product representation
\begin{equation}
\mathcal{H}_{\loc}=\mathcal{H}_{\mathrm{irrep}}^{(1/2)}\otimes \mathcal{H}_{\mathrm{irrep}}^{(1)}\simeq\mathcal{H}_{\mathrm{irrep}}^{(1/2)}\oplus \mathcal{H}_{\mathrm{irrep}}^{(3/2)}\,.
\end{equation}
Thus, the only non-vanishing multiplicities are $a_{1/2}=a_{3/2}=1$, and $k=6$.

The local partition function entering the saddle-point calculation is
\begin{equation}
Z_{\loc}(\beta)=\Big(2\cosh\frac{\beta}{2}\Big)\big(1+2\cosh\beta\big)\,.
\end{equation}
The saddle-point condition therefore reads
\begin{equation}
s=-\frac{\partial}{\partial\beta}\log Z_{\loc}(\beta)\bigg|_{\beta=\beta_*(s)}=-\frac{1}{2}\tanh\frac{\beta_*(s)}{2}-\frac{2\sinh\beta_*(s)}{1+2\cosh\beta_*(s)}\,.
\end{equation}
It is convenient to introduce
\begin{equation}
x(s)=-\tanh\frac{\beta_*(s)}{2}\,,\qquad 0<x<1\,,
\end{equation}
in terms of which the saddle-point equation becomes
\begin{equation}
s=\frac{x}{2}+\frac{4x}{3+x^2}=\frac{x(x^2+11)}{2(x^2+3)}\,,\qquad 0<s<\frac{3}{2}\,,
\label{eq:mixed-spin-saddle}
\end{equation}
or, equivalently, $x^3-2s x^2+11x-6s=0$. Equation \eqref{eq:mixed-spin-saddle} has a unique solution $x=x(s)$ in the interval $0<x<1$, and
\begin{equation}
\beta_*(s)=\log\!\left(\frac{1-x(s)}{1+x(s)}\right).
\end{equation}
Unlike in the preceding examples, the saddle-point condition \eqref{eq:mixed-spin-saddle} is now a cubic equation. Its solution, which has a closed form not reported here, is denoted $x(s)$ and kept impicit.

Because the charge generator on the local Hilbert space is the sum of the spin-$1/2$ and spin-$1$ generators, the local thermal probability distribution factorizes in the product basis $|m_{1/2}\rangle\otimes|m_1\rangle$, even though this basis does not coincide with the one in which $\mathcal{H}_{\loc}$ decomposes into $SU(2)$ irreducible representations, $j=1/2\oplus 3/2$. In terms of $x=x(s)$, its two factors are
\begin{equation}
p^{(1/2)}_{\pm 1/2}=\frac{1\pm x}{2}\,,
\end{equation}
and
\begin{equation}
p^{(1)}_{+1}=\frac{(1+x)^2}{3+x^2}\,,\qquad p^{(1)}_{0}=\frac{1-x^2}{3+x^2}\,,\qquad p^{(1)}_{-1}=\frac{(1-x)^2}{3+x^2}\,.
\end{equation}
Consequently, the local entropy is
\begin{align}
\eta_{\frac{1}{2}\otimes 1}(s)
={}&
-\sum_{m=\pm 1/2}p^{(1/2)}_m\log p^{(1/2)}_m-\sum_{m=-1}^{1}p^{(1)}_m\log p^{(1)}_m
\nonumber\\
={}&
\log\!\left(\frac{2(3+x(s)^2)}{(1-x(s)^2)^{3/2}}\right)-2s\,\mathrm{arctanh}\!\left(x(s)\right).
\label{eq:eta-SU2-mixed-spin}
\end{align}
Equivalently, this is the sum of the spin-$1/2$ and spin-$1$ thermal entropies evaluated at the same inverse temperature. Their mean charges, however, are respectively
\begin{equation}
s_{1/2}=\frac{x}{2}\,,\qquad s_1=\frac{4x}{3+x^2}\,,\qquad s=s_{1/2}+s_1\,,
\end{equation}
and are therefore not equal fractions of the total charge.

The group-dependent prefactor and local heat capacity are
\begin{equation}
\alpha_0(s)=1-\ee^{\beta_*(s)}=\frac{2x(s)}{1+x(s)}\,,
\end{equation}
and
\begin{equation}
c_*(s)=(1-x^2)\left[\frac{1}{4}+\frac{2(3-x^2)}{(3+x^2)^2}\right]\left(\log\frac{1-x}{1+x}\right)^2,\qquad x=x(s)\,.
\label{eq:beta-c-SU2-mixed-spin}
\end{equation}
The average entanglement entropy is then obtained by substituting \eqref{eq:eta-SU2-mixed-spin}, $\beta_*(s)$ and $c_*(s)$ into \eqref{eq:entropyasymptotic-SU2}.

\section{Discussion}
\label{sec:discussion}
The analysis presented in this paper provides a general treatment of typical entanglement entropy for any many-body system with a fixed global charge and a general local Hilbert space $\mathcal{H}_\loc$, in the thermodynamic limit $N\to\infty$. The known results on typical entanglement entropy $\langle S_A\rangle_{Ns}$ with fixed abelian and non-abelian charge are recovered as special cases of the general formula \eqref{eq:entropyasymptotic}:
\begin{itemize}[leftmargin=1em,label=\raisebox{.14em}{$\scriptstyle\bullet$}]
\item In the case of fixed abelian $U(1)$ charge, the average entanglement entropy \eqref{eq:entropyasymptotic} reduces to the expression \eqref{eq:entropy-U1} found in \cite{Yauk:2023wbu} for $k$-dimensional local Hilbert spaces with particle-number conservation, and in \cite{Bianchi:2021aui} for the special case of systems of qubits.
\item In the case of fixed non-abelian $SU(2)$ charge, the average entanglement entropy  \eqref{eq:entropyasymptotic} reduces to the expression \eqref{eq:entropyasymptotic-SU2} for general $k$-dimensional local Hilbert spaces, generalizing the result of \cite{Bianchi:2024aim,Chakraborty:2025ziy} for the special case of systems of qubits.
\end{itemize}
The construction presented here applies to any many-body system with a local single-charge symmetry group and a fixed global charge. We showed that the typical entanglement entropy admits a thermodynamic interpretation in terms of the local entropy $\eta(s)$ defined in \eqref{eq:eta-s-def}, where the charge density $s$ plays the role of the energy density of the local system. The function $\eta(s)$ is fully determined by the structure of the local Hilbert space $\mathcal{H}_\loc$ and its decomposition into irreducible representations of the local symmetry group. It is given by the expression $\eta(s)=\log(k)-\mathcal{D}(p_s\|p_\circledast)$, \eqref{eq:eta-s-def},
where $k$ is the dimension of $\mathcal{H}_\loc$, the probability distribution $p_s$ is the thermal distribution at fixed average charge $s$, the probability distribution $p_\circledast$ is the thermal distribution at infinite temperature, and $\mathcal{D}(p_s\|p_\circledast)$ is their relative entropy, or Kullback-Leibler divergence. When all local multiplicities are trivial ($a_{q_\loc}=1$), the infinite-temperature distribution $p_\circledast = 1/k$ is uniform, and $\eta(s)$ reduces to the Shannon entropy of $p_s$. To our knowledge, the formula \eqref{eq:eta-s-def} and the thermodynamic interpretation of $\eta(s)$ are new.

\medskip

Starting from the thermodynamic interpretation of the charge density $s$ as local energy and the function $\eta(s)$ as local entropy, we defined the inverse temperature $\beta_*(s)=\eta'(s)$ and the local heat capacity $c_*(s)=\partial s/\partial(\beta_*^{-1})$. The infinite-temperature limit $\beta_*(s_\circledast)=0$ defines the critical charge density $s_\circledast$. With these definitions, we can characterize the behavior of the average entanglement entropy $\langle S_A\rangle_{Ns}$ in thermodynamic terms:

\begin{itemize}[leftmargin=1.2em,label=\raisebox{.14em}{$\scriptstyle\bullet$}]
\item At the leading order $\O(N)$ and up to half-system size $f\leq 1/2$, the average entanglement entropy is extensive, i.e., it is proportional to the number $N_A$ of bodies in the subsystem $A$. The coefficient of the leading order term is the local entropy $\eta(s)$ defined in \eqref{eq:eta-s-def} and described above, \eqref{eq:eta-intro}. In the case of local qubit Hilbert space ($k=2$), we find that the expressions \eqref{eq:eta-U1-qubit} and \eqref{eq:eta-SU2-qubit} for fixed global $U(1)$ charge $m$ (with $s=m/N$) and global $SU(2)$ charge $j$ (with $s=j/N$) have the same form $\eta_{U(1)-\mathrm{qubit}}(s)=\eta_{SU(2)-\mathrm{qubit}}(s)$ and match the expressions found in \cite{Bianchi:2019stn,Bianchi:2021aui,Bianchi:2024aim,Vidmar:2017pak,Murthy:2019qvb,Murciano:2022lsw,Lau:2022hvc,Kliczkowski:2023qmp,Swietek:2023fka,Rodriguez-Nieva:2023err,Yauk:2023wbu,Jonay:2022cwg,Langlett:2024sxk,Ghosh:2024rvs,Langlett:2025fam,Medos:2026pkq,Patil:2023wdw,Patil:2025ump,Chakraborty:2025ziy,Wu:2026jxe,Yauk:2026quy}. In the case of local qutrit Hilbert space ($k=3$), we find again that the expressions \eqref{eq:eta-U1-qutrit} for $U(1)$ and \eqref{eq:eta-SU2-qutrit} for $SU(2)$ have the same form $\eta_{U(1)-\mathrm{qutrit}}(s)=\eta_{SU(2)-\mathrm{qutrit}}(s)$ and match the expressions found in \cite{Medos:2026pkq,Yauk:2023wbu}. We also determined the local entropy $\eta(s)$ for two further cases in which the local charge multiplicities are non-trivial. In the case of $N$ qutrits with non-uniform $U(1)$ charge (two-species hardcore bosons), we find that the local entropy $\eta_{{U(1)}\text{-2bosons}}(s)$ in \eqref{eq:eta-U1-2bosons} matches the expressions found in \cite{Yauk:2023wbu}. In the case of $N$ trimers of spin-$\frac{1}{2}$ with fixed $SU(2)$ charge, the local entropy takes the form $\eta_{SU(2)\text{-trimer}}(s)=3\,\eta_{SU(2)\text{-qubit}}(s/3)$ in \eqref{eq:eta-SU2-trimer}, showing that this system is equivalent to three independent spin-$\frac{1}{2}$ bodies grouped into a single local $SU(2)$ representation which is reducible. 

\item At the leading order $\O(N)$ and for larger subsystem fraction, $f>1/2$, the average entanglement entropy is Page-like \cite{Page:1993df}, i.e., it can be obtained by the substitution $f\leftrightarrow 1-f$. 

\item At half-system size, $f=1/2$, a term of order $\mathcal{O}(\sqrt{N})$ appears, whose coefficient is proportional to the square root of the local heat capacity \eqref{eq:c-def}. In the $U(1)$ case, this correction was first identified analytically and observed numerically in \cite{Vidmar:2017pak}, then explained in terms of energy conservation in \cite{Murthy:2019qvb} and derived from number conservation in \cite{Bianchi:2021aui}. This term vanishes in the infinite temperature limit $s\to s_\circledast$. We showed that the result is general and computed the local heat capacity $c_*(s)$ in terms of the local entropy $\eta(s)$ in each of the three examples for $U(1)$ charge \eqref{eq:beta-c-U1-qubit}, \eqref{eq:beta-c-U1-qutrit}, \eqref{eq:beta-c-U1-boson}, and the four examples for $SU(2)$ charge \eqref{eq:beta-c-SU2-qubit}, \eqref{eq:beta-c-SU2-qutrit}, \eqref{eq:beta-c-SU2-trimer}, \eqref{eq:beta-c-SU2-mixed-spin}.

\item At the order $\O(N^0)$ and generic subsystem fraction $f$, the universal term $\frac{\log(1-f) +f}{2}$ appears, which was identified in \cite{Vidmar:2017pak} for eigenstates of a physical Hamiltonian with particle number conservation. This term was shown to be universal for random states with $U(1)$ charge in \cite{Yauk:2023wbu}, and is now also shown to be universal for random states with $SU(2)$ charge, generalizing the result of \cite{Bianchi:2024aim} for spin-$1/2$ systems with a fixed $SU(2)$ charge (see \cite{Bianchi:2024aim} and \cite{Chakraborty:2025ziy} for a discussion of the relation between $K$-local and $G$-local subsystems). The terms that depend on $\alpha_0(s)$ are trivial for $U(1)$ charge, where $\alpha_0(s)=1$, but non-trivial for $SU(2)$ charge, where $\alpha_0(s)=1-\ee^{\beta_*(s)}$ \eqref{eq:alpha0-SU2}. Remarkably, in the case of the non-abelian group $SU(2)$, they result in an asymmetry under $f\leftrightarrow 1-f$, as identified in \cite{Bianchi:2024aim}. 

\item At the order $\O(N^0)$ and half-system fraction $f=1/2$, an additional term proportional to $\delta_{s,s_\circledast}$ appears. This term was found earlier for $U(1)$ charge \cite{Bianchi:2021aui,Yauk:2023wbu}, and appears only at the infinite-temperature charge density $s=s_\circledast$, defined by $\beta_*(s_\circledast)=0$. As noticed in \cite{Yauk:2023wbu}, the $\delta_{s,s_\circledast}$ term and the $\O(\sqrt{N})$ term are mutually exclusive. In the $SU(2)$ case, the value $s=s_\circledast$ is extremal and is not considerd here (See Sec.~6.2 of \cite{Bianchi:2024aim}). 

\item The variance of the entanglement entropy \eqref{eq:variance-general} is exponentially suppressed in $N$ for any charge density $s$ with non-vanishing local entropy, $\eta(s)\neq 0$. Therefore, in the thermodynamic limit, the average entanglement entropy $\langle S_A\rangle_{Ns}$ is \emph{typical}: the probability distribution $P(S_A)$ of the entanglement entropy of a random state with fixed charge $q=Ns$ is sharply peaked around the average, with fluctuations that are exponentially small in the system size \cite{Bianchi:2019stn}.
\end{itemize}

\noindent The results presented here apply to random pure states, sampled uniformly in a Hilbert space sector of fixed global charge. The construction does not rely on a specific Hamiltonian. Energy eigenstates of physical Hamiltonians also sample this same Hilbert space sector, provided that they have the same abelian or non-abelian symmetries and the same local Hilbert space structure as discussed here. We expect that the results we presented for the typical entanglement entropy can serve as a probe of quantum chaos in physical Hamiltonians as discussed in \cite{Vidmar:2017pak,Bianchi:2021aui,Medos:2026pkq}. Specifically, it would be interesting to investigate the following cases for non-abelian $SU(2)$ charge:
\begin{itemize}[leftmargin=1.2em,label=\raisebox{.14em}{$\scriptstyle\bullet$}]
\item The spin-$1/2$ Heisenberg chain \cite{Patil:2023wdw,Patil:2025ump,Chakraborty:2025ziy,Wu:2026jxe},
\begin{equation}
H=-\sum_{n=1}^N \vec{S}_n\cdot \vec{S}_{n+1}-\lambda\sum_{n=1}^N  \vec{S}_n\cdot \vec{S}_{n+2}\,,
\end{equation}
is integrable for $\lambda=0$, quantum chaotic (non-integrable) for $\lambda\neq0$, and maximally chaotic for $\lambda=3$ (see App.~C of \cite{Patil:2023wdw}). This many-body system has the same local Hilbert space as the one studied in Sec.~\ref{sec:qubit-SU2}, with energy eigenstates of fixed global $SU(2)$ charge $j$ which sample the Hilbert space \eqref{eq:HNq}. The eigenstate entanglement entropy can be compared to the typical value \eqref{eq:entropyasymptotic-SU2} as a probe of quantum chaos for different values of $\lambda$, possibly also including a spin-chirality term $\vec{S}_n\cdot (\vec{S}_{n+1}\times \vec{S}_{n+2})$, \cite{Wu:2026jxe}.

\item The spin-$1$ Heisenberg chain \cite{Patil:2026rst},
\begin{equation}
H=-\sum_{n=1}^N \vec{L}_n\cdot \vec{L}_{n+1}-\mu\sum_{n=1}^N  \vec{L}_n\cdot \vec{L}_{n+2}\,,
\end{equation}
is quantum chaotic even for $\mu=0$, and integrable for $\mu=1$ \cite{Zamolodchikov:1980ku,Babujian:1982ib,Bytsko:2001uh}. This many-body system has the same local Hilbert space as the one studied in Sec.~\ref{sec:qutrit-SU2}, with energy eigenstates of fixed global $SU(2)$ charge $j$ which sample the Hilbert space \eqref{eq:HNq}. The eigenstate entanglement entropy can be compared to the typical value \eqref{eq:entropyasymptotic-SU2} as a probe of quantum chaos for different values of $\mu$. Additional terms, such as the ones considered in \cite{Patil:2026rst}, which break the $SU(2)$ symmetry to a $U(1)$ subgroup with fixed global charge $m$, change the expected behavior of the entanglement entropy at the order $\O(N^0)$. The comparison to the typical entropy \eqref{eq:entropy-U1} in the $U(1)$ case has been studied in \cite{Vidmar:2017pak,Bianchi:2021aui}.

\item The spin-$j_\loc$ case appears in loop quantum gravity \cite{Rovelli:2014ssa,Ashtekar:2021kfp} where a Hamiltonian of the form
\begin{equation}
H=\sum_{n,m=1}^{N}c_{nm} \vec{J}_n\cdot\vec{J}_m\;+\;\sum_{n,m,r=1}^{N}g_{nmr}\, \vec{J}_n\cdot(\vec{J}_m\times  \vec{J}_r)
\end{equation}
describes the quantum geometry of a polyhedron \cite{Bianchi:2010gc,Bianchi:2011ub}. It would be interesting to investigate the behavior of the entanglement entropy in the semiclassical limit where the area of each face of the polyhedron is large, $j_\loc\gg 1$, and the fixed $SU(2)$ charge $j\gg j_\loc$ determines the base of a dome, corresponding to the finite temperature regime $\beta_*(s)\neq 0$ where $s=j/N$.

\end{itemize}
\noindent The framework introduced in this paper relies only on the structure of the local Hilbert space and the global charge. It provides a thermodynamic interpretation for the scaling of the typical entanglement entropy which we expect to apply to a broad class of physical systems with the same structure.

\section*{Acknowledgments}
This work was made possible through the support of the ID\# 63683 grant from the John Templeton Foundation, as part of the project \href{https://withoutspacetime.org/}{``WithOut SpaceTime'' (WOST)}. The opinions expressed in this work are those of the authors and do not necessarily reflect the views of the John Templeton Foundation. E.B. acknowledges support by the National Science Foundation, Grants No. PHY-2207851 and PHY-2513194.

\begin{appendix}
\section{Laplace approximation and discontinuities} 
We report here the relevant formula for the asymptotic evaluation, for $N\gg 1$, of integrals of the form
\begin{equation}
\label{eq:Laplace-continuous}
\int_{t_1}^{t_2}h(t) \,\ee^{N g(t)}\,dt\;=\;\Big(1+\frac{C_1}{N}+\O(N^{-2})\Big)\sqrt{\frac{2\pi}{-g''(t_0)\,N\,}}\; h(t_0)\,\ee^{N g(t_0)} \, ,
\end{equation}
using the Laplace method. This is a standard textbook treatment, but it is difficult to find all the formulas we need in one place in the literature. Here we assume that the functions $g(t)$ and $h(t)$ are smooth, and there is a single maximum of $g(t)$ at the point $t_0$ in the interval $(t_1,t_2)$, i.e.,
\begin{equation}
g'(t_0)=0\,,\quad g''(t_0)<0\,.
\end{equation}
Since $g(t)<g(t_0)$ for all $t\neq t_0$ in $(t_1,t_2)$, the integral is exponentially concentrated near $t_0$. Expanding both functions around $t_0$,
\begin{equation}
  g(t)=g(t_0)+\frac{1}{2}g''(t_0)(t-t_0)^2+O((t-t_0)^3) \, , \qquad
  h(t)=h(t_0) + O(t-t_0)\, ,
\end{equation}
the integral becomes
\begin{equation*}
\int_{t_1}^{t_2}h(t)\,\ee^{Ng(t)}\,dt
\;=\;h(t_0)\,\ee^{Ng(t_0)}\int_{t_1}^{t_2}\ee^{\frac{N}{2}g''(t_0)(t-t_0)^2}\,\bigl(1+\O(N^{-1})\bigr)\,dt\,.
\end{equation*}
Changing variables to $\tau=t-t_0$ and extending the limits to $\pm\infty$ at exponentially small cost, the remaining integral is a standard Gaussian, giving the leading term in \eqref{eq:Laplace-continuous}:
\begin{equation*}
\int_{t_1}^{t_2}h(t)\,\ee^{Ng(t)}\,dt
\;=\;\bigl(1+\O(N^{-1})\bigr)\sqrt{\frac{2\pi}{-g''(t_0)\,N}}\,h(t_0)\,\ee^{Ng(t_0)}\,.
\end{equation*}
To compute the correction $C_1$, we retain higher-order terms in the Taylor expansions and integrate them against the Gaussian weight. The Gaussian moments vanish for odd $k$ by symmetry, and for even $k$ evaluate to
\begin{equation*}
\overline{(t-t_0)^k}\;\equiv\;\int_{-\infty}^{\infty} \tau^k\,\ee^{\frac{N}{2}g''(t_0)\tau^2}\,d\tau\;=\;\frac{(k-1)!!}{\bigl(-g''(t_0)\bigr)^{k/2}}\frac{1}{N^{(k+1)/2}}\sqrt{\frac{2\pi}{-g''(t_0)}}
 \,.
\end{equation*}
We factorize the Gaussian factor in the integrand  
\begin{equation*}
  h(t)\,\ee^{Ng(t)}=\bigl(h(t)\,\ee^{Ng(t)-\frac{N}{2}g''(t_0)(t-t_0)^2}\bigr)\;\ee^{\frac{N}{2}g''(t_0)(t-t_0)^2}
\end{equation*}
and Taylor-expand the prefactor. A term of the form $N^{k_1}(t-t_0)^{k_2}$, integrated against the Gaussian, contributes at order $\O(N^{k_1-(k_2+1)/2})$ for even $k_2$. The contributions at relative order $\O(N^{-1})$, i.e., those with $k_2=2(k_1+1)$, are:
\begin{align*}
\frac{1}{2}h''(t_0)(t-t_0)^2  &\quad(k_1=0, k_2=2)\,,\\
\frac{N}{6} h'(t_0)g'''(t_0) (t-t_0)^4  &\quad(k_1=1, k_2=4)\,,\\
\frac{N}{24} h(t_0) g''''(t_0)(t-t_0)^4 &\quad(k_1=1, k_2=4)\,,\\
\frac{N^2}{36} h(t_0) (g'''(t_0))^2 (t-t_0)^6 &\quad(k_1=2, k_2=6)\,.
\end{align*}
Evaluating each contribution and normalizing by the leading result gives $C_1$:
\begin{equation}
\textstyle C_1\,=\,-\frac{h''(t_0)}{2\,h(t_0)\,g''(t_0)}+\frac{h'(t_0)\,g'''(t_0)}{2\,h(t_0)\,g''(t_0)^2}-\frac{5\,g'''(t_0)^2}{24\,g''(t_0)^3}+\frac{g''''(t_0)}{8\,g''(t_0)^2}\,.
\end{equation}

This result can be generalized to the case where the function $h(t)$ is not smooth. Consider $h(t)$ discontinuous, with discontinuous derivatives, at the maximum $t_0$, i.e.,
\begin{equation}
h(t)=
\begin{cases}
\;h_-(t)\quad&t<t_0\,,\\[.2em]
\;h_+(t)&t>t_0\,.
\end{cases}
\end{equation}
The main difference from the smooth case is that the leading term is proportional to the average $\frac{h_-(t_0)+h_+(t_0)}{2}$ of the one-sided limits at $t_0$, rather than $h(t_0)$, which is no longer well defined. Moreover, the discontinuity introduces half-integer power corrections in $N$, absent in the smooth case.
\begin{equation}
\label{eq:Laplace-discontinuous}
\int_{t_1}^{t_2}h(t) \,\ee^{N g(t)}dt\;=\Big(1+\tfrac{\tilde{C}_{1/2}}{\sqrt{N}}+\tfrac{\tilde{C}_{1\vphantom{1/2}}}{N\vphantom{\sqrt{N}}}+\O(N^{-3/2})\Big)\sqrt{\tfrac{2\pi}{-g''(t_0)\,N\,}}\; \tfrac{h_-(t_0)+h_+(t_0)}{2}\,\ee^{N g(t_0)}\, .
\end{equation}

The derivation follows the same steps as the smooth case, but with the integration domain split at $t_0$ into $[t_1,t_0)$ and $(t_0,t_2]$, on each of which $h_\pm$ is smooth. The discontinuity of $h(t)$ at $t_0$ breaks the left-right symmetry of the integrand, so odd Gaussian moments no longer cancel between the two halves. In particular, for odd $k_2$,
\begin{equation*}
\int_{-\infty}^{0}N^{k_1}\tau^{k_2}\, e^{\frac{N}{2}g''(t_0)\tau^2}\,d\tau \;=\; -\frac{(k_2-1)!!}{\bigl(-g''(t_0)\bigr)^{(k_2+1)/2} } N^{k_1-(k_2+1)/2}\, , \qquad (\text{with } k_2 \text{ odd})\,,
\end{equation*}
and similarly for $(0,\infty)$, giving non-zero half-integer power corrections. The terms contributing to $\tilde{C}_{1/2}$, i.e., those with $k_2=2k_1+1$, are:
\begin{align*}
h'(t_0)(t-t_0)  &\quad(k_1=0, k_2=1)\,,\\
\frac{N}{3} h(t_0)g'''(t_0) (t-t_0)^3  &\quad(k_1=1, k_2=3)\,.
\end{align*}
Evaluating each contribution and normalizing by the leading result gives the coefficient $\tilde{C}_{1/2}$
\begin{equation}
\label{eq:C12}
\textstyle \tilde{C}_{1/2}=\frac{1}{\sqrt{-2\pi\, g''(t_0)}}\,\Big(
2\,\frac{h_{+}'(t_0) - h_{-}'(t_0)}{h_{-}(t_0) + h_{+}(t_0)}
+
\frac{2}{3}\frac{h_{-}(t_0) - h_{+}(t_0)}{h_{-}(t_0) + h_{+}(t_0)}\,
\frac{g'''(t_0)}{g''(t_0)}\Big)\,.\\[.5em]
\end{equation}
The terms contributing to $\tilde{C}_1$ satisfy $k_2=2(k_1+1)$, identical to the smooth case, but with $h(t_0)$, $h'(t_0)$ and $h''(t_0)$ replaced by their averages across the discontinuity:
\begin{equation}
\textstyle \tilde{C}_{1}=
-\frac{h_{-}''(t_0) + h_{+}''(t_0)}{2\,(h_{-}(t_0) + h_{+}(t_0))\,g''(t_0)}
+\frac{\big(h_{-}'(t_0) + h_{+}'(t_0)\big)\,g'''(t_0)}{2\,(h_{-}(t_0) + h_{+}(t_0))\,g''(t_0)^2}
-\frac{5\,g'''(t_0)^2}{24\,g''(t_0)^3}
+\frac{g''''(t_0)}{8\,g''(t_0)^2}\,.
\end{equation}



\subsection{$Y_3$ term}
\label{app:Y3}
We report here the detailed calculation of the term $Y_3$ defined in \eqref{eq:Y3}. Using the definitions of $d_{q_A}$, $b_{q,q_A}$, and $D_q$, we can write $Y_3$ as
\begin{equation}
    Y_3=-\frac{1}{2}\sum_{q_A}\frac{\min{(d_{q_A}^2},b_{q,q_A}^2)}{D_q}=-\frac{1}{2}\int_a^b\frac{\min{(d(t)^2,b(s,t)^2)}}{D(s)} f N dt + \O(N^{-1})\,.
\end{equation}
We now identify the conditions for which $Y_3$ contributes at order $O(N^0)$. The function $\min(d(t)^2,b(s,t)^2)$ has a discontinuous derivative at the point $t_{\text{crit}}$ defined by $d(t_{\text{crit}})=b(s,t_{\text{crit}})$, so the integral must be split into two integrals, each with a smooth integrand,
\begin{equation}
    Y_3=\sqrt{N}\int_a^{t_{\text{crit}}}h_-(t)\ee^{N g_-(t)}dt+\sqrt{N}\int^b_{t_{\text{crit}}}h_+(t)\ee^{N g_+(t)}dt+\O(N^{-1})\,.
\end{equation}
If the maximum $s_\circledast$ of the exponent function 
\begin{equation}
    g(t)=
\begin{cases}
g_-(t) & t \le t_{\text{crit}} \, ,\\
g_+(t) & t > t_{\text{crit}} \, ,
\end{cases}
\end{equation}
lies at a point where $g(t)$ is smooth, i.e., for $s_\circledast\neq t_{\text{crit}}$, we can apply the Laplace method \eqref{eq:Laplace-continuous} at leading order and obtain either 
\begin{equation}
    Y_3=\frac{\alpha_0(s_\circledast)^2}{\alpha_0(s)} 
\sqrt{\frac{\eta''(s_\circledast)}{8 f\, \eta''(s)}} 
\, e^{-N \left(\eta(s) - 2 f\, \eta(s_\circledast)\right)}+\dots \, ,
\end{equation}
or 
\begin{equation}
    Y_3=\frac{1}{\alpha_0(s)}
\sqrt{
\frac{\eta''\left(\frac{s - f\, s_{\circledast}}{1 - f}\right)}
{8 (1 - f)\, \eta''(s)}
}
\, e^{-N \left(\eta(s) + 2 (f - 1)\, \eta\left(\frac{s - f\, s_{\circledast}}{1 - f}\right)\right)}+\dots \, ,
\end{equation}
This shows that $Y_3$ is exponentially suppressed unless $s_\circledast=t_{\text{crit}}$, fixing the parameters $f$ and $s$ for which $Y_3$ is of order $\O(N^0)$. From $Y_2$, where the term scaling as $\sqrt{N}$ was also found for $d(t_{\text{crit}})=b(s,t_{\text{crit}})$, we know that this condition is only satisfied at $f=\frac{1}{2}$. Using this fact we find the relation between $s$ and $s_\circledast$: 
\begin{equation*}\textstyle
\log{\left(\frac{d(t_{\text{crit}})}{b(s,t_{\text{crit}})}\right)} \,N^{-1}=f\,\eta(t_\text{crit})-(1-f)\eta\left(\tfrac{s - f\, t_{\text{crit}}}{1 - f}\right)+\O(N^{-1})=\frac{\,\eta(s_\circledast)-\eta\left(2s-s_\circledast\right)}{2}+\O(N^{-1})\, .
\end{equation*}
Since this leading term vanishes by the definition of $t_{\text{crit}}$, we obtain that $Y_3$ is exponentially suppressed unless $f=\frac{1}{2}$ and $s=s_\circledast$. Imposing these conditions we obtain 
\begin{equation}
\label{eq:gplusminus}
  g_{-}(t) = \eta(t) -\, \eta(s_\circledast) \, , \quad \text{and} \quad
    g_{+}(t) =   \eta\!\left(2s_\circledast- t\right)-\eta(s_\circledast) \, .
\end{equation}
Now, since the maximum $s_{\circledast}$ lies at a limit of integration, it is not necessarily a stationary point; i.e., we may have $g'_{\pm}(s_{\circledast}) \neq 0$. However, if that is the case, since $g_{\pm}(s_\circledast)=0$, the integrals would be of order $\O(N^{-1/2})$. The first one would be
\begin{equation}
    \sqrt{N}\int_a^{s_\circledast}h_-(t)\ee^{N g_-(t)}dt=\frac{h_-(s_\circledast)}{g_-'(s_\circledast)\,\sqrt{N}}+\O(N^{-1})\, ,
\end{equation}
and similarly the second one.

Therefore an $\O(N^0)$ contribution requires $g_\pm'(s_\circledast)=0$, which is satisfied for $f=\frac{1}{2}$ and $s=s_\circledast$ with $\eta'(s_\circledast)=0$. In that case,
\begin{equation}
    \label{eq:Laplace-half-Gaussian}
    Y_3=\sqrt{\frac{\pi}{-2g_-''(s_\circledast)}}h_-(s_\circledast)+\sqrt{\frac{\pi}{-2g_+''(s_\circledast)}}h_+(s_\circledast)+\O(\tfrac{1}{\sqrt{N}})\, .
\end{equation}
Evaluating the functions $g_\pm$ from \eqref{eq:gplusminus} and substituting the explicit expressions for $h_\pm(t)$, we find
\begin{equation}
    h_{-}(t) = \frac{\alpha_0(t)^2 \, \eta''(t)}
{2 \sqrt{2\pi}\, \alpha_0(s_\circledast)\, \sqrt{-\eta''(s_\circledast)}}\, ,
\qquad \text{and} \qquad
    h_{+}(t) =
\frac{\eta''\left(2s_\circledast- t\right)}
{2\sqrt{2\pi}\, \alpha_0(s_\circledast)\, \sqrt{-\eta''(s_\circledast)}}\, .
\end{equation}
Substituting into \eqref{eq:Laplace-half-Gaussian}, we obtain
\begin{equation}
    Y_3=-\frac{\alpha_0(s_\circledast)+\alpha_0(s_\circledast)^{-1}}{4}\delta_{f,\frac{1}{2}}\delta_{s,s_\circledast}+\O(\tfrac{1}{\sqrt{N}})\,.
\end{equation}

\end{appendix}

\bibliographystyle{JHEP}
\bibliography{biblio}

\end{document}